\documentclass[twocolumn, tighten, astrosymb]{aastex632}
\pdfoutput=1 

\usepackage{amsmath,amstext}
\usepackage{graphicx}

\newcommand{\aox}{\ifmmode{\alpha_{\mathrm{ox}}} \else $\alpha_{\mathrm{ox}}$\fi} 
\newcommand{\atoms}{\ifmmode{\mathrm{\,atoms~cm^{-2}}} \else \,atoms cm$^{-2}$\fi}
\newcommand{\ax}{\ifmmode{\alpha_x} \else $\alpha_x$\fi} 
\newcommand{\bprp}{\ifmmode{G_{BP}-G_{RP}} \else $G_{BP} - G_{RP}$\fi} 
\newcommand{\cmsq}{\ifmmode{\mathrm{cm^{-2}}} \else cm$^{-2}$\fi}
\newcommand{\degsq}{\ifmmode {\mathrm{deg^2}} \else deg$^2$\fi}
\newcommand{\perdegsq}{\ifmmode {\mathrm{deg^{-2}}} \else deg$^{-2}$\fi}
\newcommand{\ew}{\ifmmode{W_{\lambda}} \else $W_{\lambda}$\fi}
\newcommand{\fbol}{\ifmmode f_{\mathrm{bol}} \else $f_{\mathrm{bol}}$\fi} 
\newcommand{\fcgs}{\ifmmode \mathrm{erg~cm^{-2}~s^{-1}}\else erg~cm$^{-2}$~s$^{-1}$\fi}
\newcommand{\flamcgs}{\ifmmode \mathrm{erg\,cm^{-2}\,s^{-1}\,\AA^{-1}}\else erg\,cm$^{-2}$\,s$^{-1}$\,\AA$^{-1}$)\fi}
\newcommand{\fnucgs}{\ifmmode {\mathrm{erg~cm^{-2}~s^{-1}~Hz^{-1}}}\else erg~cm$^{-2}$~s$^{-1}$~Hz$^{-1}$\fi}
\newcommand{\gax }{{\lower0.8ex\hbox{$\buildrel >\over\sim$}}}
\newcommand\Ha{\ifmmode {\mathrm H}\alpha \else H$\alpha$\fi}
\newcommand\Hb{\ifmmode {\mathrm H}\beta \else H$\beta$\fi}
\newcommand{\kms}{\ifmmode~{\mathrm{km~s}}^{-1}\else ~km~s$^{-1}~$\fi}
\newcommand{\lax }{{\lower0.8ex\hbox{$\buildrel <\over\sim$}}}
\newcommand{\lcgs}{\ifmmode \mathrm{erg~s^{-1}}\else erg~s$^{-1}$\fi}
\newcommand{\lnucgs}{\ifmmode erg~s^{-1}~Hz^{-1}\else erg~s$^{-1}$~Hz$^{-1}$\fi}

\newcommand{\logz}{\ifmmode{\mathrm{log}}~z \else log$~z$\fi}
\newcommand{\lo}{\ifmmode l_o \else $~l_o$\fi}
\newcommand{\Lo}{\ifmmode L_o \else $~L_o$\fi}
\newcommand{\lx}{\ifmmode l_x \else $~l_x$\fi}
\newcommand{\Lx}{\ifmmode L_x \else $~L_x$\fi}
\newcommand{\lbol}{\ifmmode L_{\mathrm{bol}} \else $L_{\mathrm{bol}}$\fi}
\newcommand{\Lbol}{\ifmmode L_{\mathrm{bol}} \else $L_{\mathrm{bol}}$\fi}
\newcommand{\LBol}{\ifmmode L_{\mathrm{bol}} \else $L_{\mathrm{bol}}$\fi}
\newcommand{\LEdd}{\ifmmode L_{\mathrm{Edd}} \else $L_{\mathrm{Edd}}$\fi}
\newcommand{\LxLbol}{\ifmmode L_x/L_{\mathrm{bol}} \else $L_x/L_{\mathrm{bol}}$\fi}
\newcommand{\rEdd}{\ifmmode L/L_{\mathrm{Edd}} \else $L/L_{\mathrm{Edd}}$\fi}
\newcommand{\REdd}{\ifmmode L/L_{\mathrm{Edd}} \else $L/L_{\mathrm{Edd}}$\fi}
\newcommand{\Rblr}{\ifmmode {R_{\mathrm BLR}} \else $R_{\mathrm BLR}$\fi}
\newcommand{\lamEdd}{\ifmmode \lambda_{\mathrm{Edd}} \else $\lambda_{\mathrm{Edd}}$\fi}
\newcommand{\mbh}{\ifmmode {M_{\rm BH}}\else${M_{\rm BH}}$\fi}
\newcommand{\Mbh}{\ifmmode {M_{\rm BH}}\else${M_{\rm BH}}$\fi}
\newcommand{\mdot}{\ifmmode \dot{m} \else $\dot{m}$\fi}
\newcommand{\mdote}{\ifmmode \dot{m}_{E} \else $\dot{m}_{E}$\fi}
\newcommand{\mone}{\ifmmode ^{-1}\else$^{-1}$\fi}
\newcommand{\msun}{\ifmmode {M_{\odot}}\else${M_{\odot}}$\fi}
\newcommand{\Msun}{\ifmmode {M_{\odot}}\else${M_{\odot}}$\fi}
\newcommand{\mtwo}{\ifmmode ^{-2}\else$^{-2}$\fi}
\newcommand{\Mvir}{\ifmmode {M_{\rm BH}^{\mathrm SE}}\else${M_{\rm BH}^{\mathrm SE}}$\fi}
\newcommand{\nhgal}{\ifmmode{ N_{H}^{Gal}} \else N$_{H}^{Gal}$\fi}
\newcommand{\nh}{\ifmmode{\mathrm N_{H}} \else N$_{H}$\fi}
\newcommand{\nhintr}{\ifmmode{ N_{H}^{intr}} \else N$_{H}^{intr}$\fi}
\newcommand{\nhtot}{\ifmmode{ N_{H}^{tot}} \else N$_{H}^{tot}$\fi}
\newcommand{\nhz}{\ifmmode{ N_{H}^z} \else N$_{H}^z$\fi}
\newcommand{\oi}{\ifmmode{\mathrm [O\,II]} \else [O\,II]\fi}
\newcommand{\oii}{\ifmmode{\mathrm [O\,II]} \else [O\,II]\fi}
\newcommand{\oiii}{\ifmmode{\mathrm [O\,III]} \else [O\,III]\fi}
\newcommand{\optebl}{\ifmmode L_{\rm 2500\,\AA} \else $~L_{\rm 2500\,\AA}$\fi}
\newcommand{\opteml}{\ifmmode l_{\mathrm{2500\,\AA}} \else $~l_{\mathrm{2500\,\AA}}$\fi}
\newcommand{\rhodC}{\ifmmode{ \rho_{\mathrm{dC}}} \else $\rho_{\mathrm{dC}}$ \fi}
\newcommand{\Teff}{\ifmmode T_{\mathrm{Eff}} \else $T_{\mathrm{Eff}}$\fi}
\newcommand{\xebl}{\ifmmode L_X \else $~L_X$\fi}
\newcommand{\xeml}{\ifmmode l_{\mathrm{2\,keV}} \else $~l_{\mathrm{2\,keV}}$\fi}

\def\geqsim{\lower.73ex\hbox{$\sim$}\llap{\raise.4ex\hbox{$>$}}$\,$}
\def\leqsim{\lower.73ex\hbox{$\sim$}\llap{\raise.4ex\hbox{$<$}}$\,$}

\newcommand{\umg}{\ifmmode{\mathrm{(}u-g\mathrm{)}} \else ($u-g$)\fi}
\newcommand{\gmr}{\ifmmode{\mathrm{(}g-r\mathrm{)}} \else ($g-r$)\fi}
\newcommand{\rmi}{\ifmmode{\mathrm{(}r-i\mathrm{)}} \else ($r-i$)\fi}
\newcommand{\gmi}{\ifmmode{\mathrm{(}g-i\mathrm{)}} \else ($g-i$)\fi}
\newcommand{\imz}{\ifmmode{\mathrm{(}i-z\mathrm{)}} \else ($i-z$)\fi}
\newcommand{\jmh}{\ifmmode{\mathrm{(}J-H\mathrm{)}} \else ($J-H$)\fi}
\newcommand{\hmk}{\ifmmode{\mathrm{(}H-K\mathrm{)}} \else ($H-K$)\fi}
\newcommand{\ctwo}{\ifmmode C_2 \else C$_2$\fi}

\received{December, 02, 2024}
\revised{January, 25, 2025}
\accepted{January, 27, 2025}

\submitjournal{\apj}

\shorttitle{Carbon Stars from Gaia DR3}
\shortauthors{Roulston et al.}

\turnoffediting

\begin{document}

\title{Carbon Stars From Gaia DR3 and the Space Density of Dwarf Carbon Stars}

\correspondingauthor{Benjamin R. Roulston}
\email{broulsto@clarkson.edu}

\author[0000-0002-9453-7735]{Benjamin R. Roulston}
\affiliation{Department of Physics, Clarkson University, 8 Clarkson Ave, Potsdam, NY 13699, USA}

\author[0000-0001-5782-3719]{Naunet Leonhardes-Barboza}
\affiliation{Department of Astronomy and Astrophysics, University of California, Santa Cruz, CA 95064, USA}
\affiliation{Center for Astrophysics $\vert$ Harvard \& Smithsonian, 60 Garden Street, Cambridge, MA 02138, USA}
\affiliation{Department of Astronomy, Wellesley College, Wellesley, MA 02841 USA}

\author[0000-0002-8179-9445]{Paul J. Green}
\affiliation{Center for Astrophysics $\vert$ Harvard \& Smithsonian, 60 Garden Street, Cambridge, MA 02138, USA}

\author{Evan Portnoi}
\affiliation{Division of Physics, Mathematics, and Astronomy, California Institute of Technology, Pasadena, CA 91125, USA}





\begin{abstract}
Carbon stars (with atmospheric C/O$>1$) range widely in temperature and luminosity, from low mass dwarfs to asymptotic giant branch stars (AGB). The main sequence dwarf carbon (dC) stars have inherited carbon-rich material from an AGB companion, which has since transitioned to a white dwarf.  The dC stars are far more common than C giants, but no reliable estimates of dC space density have been published  to date.  We present results from an all-sky survey for carbon stars using the low-resolution XP spectra from Gaia DR3. We developed and measured a set of spectral indices contrasting C$_{\rm 2}$ and CN molecular band strengths in carbon stars against common absorption features found in normal (C/O$<1$) stars such as CaI, TiO and Balmer lines. We combined these indices with the XP spectral coefficients as input to supervised machine-learning algorithms trained on a vetted sample of known C stars from LAMOST.  We describe the selection of the carbon candidate sample, and provide a catalog of 43,574 candidates dominated by cool C giants in the Magellanic Clouds and at low galactic latitude in the Milky Way. We report the confirmation of candidate C stars using intermediate ($R\sim 1800$) resolution optical spectroscopy from the Fred Lawrence Whipple Observatory, and provide estimates of sample purity and completeness.  From a carefully-vetted sample of over 600 dCs, we measure their local space density to be $\rho_0\,=\,1.96^{+0.14}_{-0.12}\times10^{-6}\,\text{pc}^{-3}$ (about one dC in every local disk volume of radius 50\,pc), with a relatively large disk scale height of $H_z\,=\,856^{+49}_{-43}\,$pc.
\end{abstract}



\section{Introduction \label{sec:intro}}

Strong atmospheric carbon (C$>$O) is only intrinsic to asymptotic giant branch (AGB) stars that have undergone the third dredge-up (thermally pulsing, or TP-AGB) phase, whereby strong convection brings helium, carbon, and $s$-process elements to the surface \citep{Iben1974, Iben1983}.  However, a whole family of carbon and related stars exist. Carbon stars are traditionally broken up into the following categories: C-H, C-R, and C-N. Other notable categories are C-J, Barium (Ba), CH, and carbon-enhanced metal-poor (CEMP) stars \citep{Wallerstein1998}. C-N, C-H, and C-R can be differentiated by their carbon band strength and colors. C-N stars are the reddest stars with the deepest carbon bands -- typically AGB stars and other giants. Since AGB C stars are luminous, they are easily detectable to large distances in the halo of the Milky Way, as well as in the Large and Small Magellanic Clouds and other nearby galaxies. C-H and C-R are bluer stars, often subgiants and difficult to distinguish from each other. These stars' carbon bands are weaker in strength than C-N stars.  Many C-R stars in the literature are actually misclassified C-N or C-H stars, but the evolution of the genuine early C-R stars is still debated \citep{Izzard2007,Zamora2009}. Barium stars tend to have relatively weak carbon bands, similar in strength to some of the early C-R stars, however, these are also accompanied by Ba II absorption lines at $4007$\,\AA\, and $4554$\,\AA.  A high binary fraction has been found for C-H, Ba and the $s-$process rich CEMP$-s$ stars \citep{McClure1990,Jorissen1998,Jorissen2016,Lucatello2005}.

The bulk of the literature on C stars concerns C giants and C-AGB stars.
Significant progress has been made in theoretical modeling of nucleosynthesis in C-AGB stars and their production of gas and dust as a function of mass and metallicity \citep{Busso1999,Karakas2014,Cristallo2016,Ventura2020}.  \citet{Claussen1987} studied an infrared-selected, flux-limited sample of C giants in the Milky Way and derived their space density.
Luminosities of different classes of C giants in the Milky Way are considered in \citet{Abia2022}, and the C-AGB luminosity function is well-studied in \citet{Straniero2023}. 

Some AGB C stars have main sequence companions that may accrete substantial C-enriched material from the C-AGB star, increasing the C/O ratio in the atmosphere of the companion. This accretion occurs primarily in the TP-AGB phase which lasts only a few million years. The AGB C star will then evolve to a white dwarf via mass-loss, which in turn cools until it becomes less luminous than its main sequence companion at optical wavelengths, leaving behind a main sequence star with a carbon-enriched atmosphere.  Depending on the mass of accreted material, its C/O ratio, the mass of the main sequence companion, its metallicity, and the depth of mixing, C$_2$ and/or CN molecular bands may become easily detectable even in medium or low-resolution spectra \citep{Dahn1977,Green1991,Christlieb2001,Li2018}. Thus, a dwarf carbon (dC) star is born.

As evidence for this scenario, in some cases the hot white dwarf is still evident in the spectrum of a DA/dC double-line spectroscopic binary \citep{Heber1993,Liebert1994,Si2014}. Also, the binary fraction of dCs appears to be near 95\% \citep{Roulston2019}. The dC stars will eventually evolve off the main sequence, and may thus be the progenitors of some of the other types of carbon giants already mentioned, at least those with high binarity such as the C-H, Ba, and  CEMP-$s$ stars (e.g., \citealt{McClure1990, Hansen2016, Izzard2010}).

The known types of carbon stars thus span all luminosities from main sequence (dwarf) to giant, to AGB (e.g., see the carbon color-magnitude diagram of \citealt{Green2019}, Figure 2).  Given that C-AGB lifetimes are typically a few Myr, $\sim 10^3 \times$ shorter than main sequence lifetimes, dC stars should be far more common in the Milky Way than C giants \citep{Kool1995}. However, though dCs have much higher space densities than C-AGB stars, they are typically 10$^{2-3}\times$ less luminous, so are less well-known from early large-area surveys with bright magnitude limits (e.g., \citealt{Stephenson1985, Sanduleak1988}). By far most dCs discovered to date are known from deeper, recent surveys like SDSS or LAMOST (e.g., \citealt{Green2013, Li2024}). However, these are subject to strong targeting and selection effects, and do not cover the whole sky. As post-mass transfer main sequence stars, dCs can have absolute magnitudes and colors ranging from those similar to M dwarfs ($M_G\sim 11$, $\bprp\sim 2.2$) up through about early G-type ($M_G\sim 4.5$, $\bprp\sim 0.7$). A C star color-magnitude diagram is shown in \citealt{Roulston2022} (Figure\,2), and example dC spectra in \citealt{Roulston2020}, (Figure\,1).   Even warmer, more massive C$>$O main sequence stars may exist, but they are likely too warm to have detectable molecular bands of C$_2$ and CN.

The brightest known dCs (defined here as having $M_G >4.5$) with spectroscopic confirmation
are about $V\sim 14$\,mag, but at most a handful are known (e.g., G77-61 and LP\,318-342). 

A new sample of bright, uniformly-selected dCs would be particularly useful, because high resolution spectroscopy could be used to validate atmospheric models and thereby key stellar parameters for dCs, such as temperature and mass. High resolution spectroscopy could also be used to measure the carbon and neutron capture elements of AGB star material still in dC atmospheres, as yet unperturbed by evolution up the giant branch.  This is of particular interest as the CH, Ba and the carbon-enhanced metal poor (CEMP-s) stars \citep{Lucatello2005} - mostly giants or subgiants - likely evolved from dC stars, and are better known than dCs from previous surveys only because of their larger luminosities. 

Products from the Gaia mission enable selection of bright C stars across the entire sky, including carbon AGB stars, red giants and dwarfs.  In this paper, we produce an all-sky catalog of C star candidates using low-resolution spectra from Gaia Data Release 3 \citep[DR3;][]{GaiaDR3} and machine learning.  While our catalog includes C stars across a wide range of luminosities, our scientific focus for this paper is on the dwarf carbon stars, their luminosity function and their space density.  Nevertheless, the C giants will prove useful for many other studies, potentially including e.g., the detection and characterization of stellar streams in the halo of the Milky Way (e.g., \citealt{Ibata2013,Chandra2023}), luminosity functions, period-luminosity relations for Mira variables, mass ranges and metallicity dependence of C-AGB star populations in the Magellanic Clouds (e.g., \citealt{Pastorelli2020,Iwanek2021}) etc.

Our primary aim is to identify a relatively pure and complete sample of dCs, to enhance the currently known sample of $\simeq$1000 dCs \citep{Li2024, Green2013},  adding new - and mostly brighter -  Gaia-identified dCs. An unbiased, all-sky sample of carbon stars will enable us to determine the dC space density using Gaia parallaxes. This is crucial for comparison to models of binary evolution leading to post-mass transfer systems. \citet{Roulston2021} found that a significant fraction of dCs are variable. Using this sample, we can also find and study new variable dCs using extant multi-epoch wide-area imaging surveys such as e.g., the Transiting Exoplanet Survey Satellite \citep[TESS;][]{Ricker2015}, the Zwicky Transient Facility \citep[ZTF;][]{Bellm2019}, the Asteroid Terrestrial-impact Last Alert System \citep[ATLAS;][]{Tonry2018}, and Gaia DR3 multi-epoch photometry \citep{GaiaDR3_multiepoch}.  A large, bright, uniformly-selected sample of dCs can help determine whether the main pathway to dC variability is from accretion spin-up, activity and spot rotation, or tidal locking and ellipsoidal modulation. Similarly, we can measure activity fraction in dCs using optical spectra and X-ray emission \citep{Roulston2022, Green2019}. A bright sample of dCs could reveal the first known eclipsing dC systems, enabling the derivation of their mass, radius and density.

In Section\,\ref{sec:GaiaDR3}, we briefly describe Gaia DR3, the parent catalog of our all-sky search for C stars.  
We describe our training set of vetted C stars and our control sample in Section\,\ref{sec:cstarTraining}.  The features we use in training (spectral indices and XP coefficients) are detailed in Section\,\ref{sec:features}, along with a discussion of optimization of the machine learning algorithms in Section\,\ref{sec:ML}.  In Section\,\ref{sec:SelectionResults} we detail the results of the machine learning classification, and  completeness calculations based on previous spectroscopic surveys. Our spectroscopic observations and measurements of sample purity are outlined in Section\,\ref{sec:FAST}.  These enable our calculation of the luminosity function and space density of dCs in Section\,\ref{sec:lfsd}.  We contrast our results to other relevant stellar samples in Section\,\ref{sec:compareSD}, and summarize our results and discuss future prospects in Section\,\ref{sec:summary}.

\section{Gaia DR3  \label{sec:GaiaDR3}}

Gaia Data Release 3 (DR3; \citealt{GaiaDR3}) has made an extensive and novel data set available, including about 220 million flux-calibrated low-resolution XP (BP/RP, Blue Photometer and Red Photometer) spectra that cover the wavelength range of $336-1020$\,nm with a spectral resolution of R\,$\sim 20 - 70$. 

Because there are no flat fields or spectroscopic lamp images taken on the BP/RP spectro-photometer on Gaia, the spectra must be internally calibrated. The internal calibration is done on a set of basis functions, leading to a set of 110 Hermite polynomial coefficients (herein referred to as coefficients) from the BP/RP spectro-photometers from which full spectrum can be reconstructed (e.g., using \texttt{GaiaXPy}; \citealt{GaiaXPy}). The photometry, $BP$ and $RP$ magnitudes, are obtained from these sampled spectra. 

It has been shown that stars with different effective temperatures show distinct distributions of the coefficients, therefore distinguishing between different stellar spectra is possible with only the coefficients \citep{Carrasco2021}. Additionally, using the coefficients can prevent the loss of information that occurs when using sampled spectra. This discrete set of 110 coefficients from Gaia DR3 is potentially powerful for classification purposes using machine learning where many features can be used as input. 

Gaia EDR3 and DR3 data have been used to calculate the luminosity functions and space densities of various classes of objects such as cataclysmic variables (CVs,) whie dwarfs (WDs), and other stars \citep{Rebassa-Mansergas2021, Rix2021, Canbay2023, Dawson2024}, adding to those from previous data releases \citep{Pala2020}. Our own measurements of the luminosity function and space density of dCs will be explained in detail in Section\, \ref{sec:lfsd}.

\section{Carbon Star Training Set}\label{sec:cstarTraining}

In order to use and test the various machine learning algorithms available, we compiled a training sample of stars with known spectral classifications. This set includes spectroscopically confirmed C stars, along with a control sample of normal (C/O $< 1$) stars.

Our first training sample was based on spectroscopically confirmed C stars from the literature, obtained through SIMBAD classifications, cross matched to those stars with XP spectra in Gaia DR3. While we were able to train, and predict with this sample, our follow-up verification spectroscopy revealed a disappointingly low purity. While we present details of the this first training sample in the appendix, we focus on our final training set and the results obtained from it.

\subsection{Carbon Star Training Sample \label{sec:TrainingSample}}

We used the Large Sky Area Multi-Object Fiber Spectroscopic Telescope (LAMOST) Galactic spectroscopy survey \citep{Zhao2012} to select our training sample of bright C stars. We selected stars from LAMOST DR8 v2.0 \citep{Deng2012} that are classified by their pipeline as C stars. This resulted in a sample of 4132 unique stars after matching to Gaia and requiring each star to have an XP spectrum available.

\begin{figure*} 
\centering
\includegraphics[width=\textwidth]{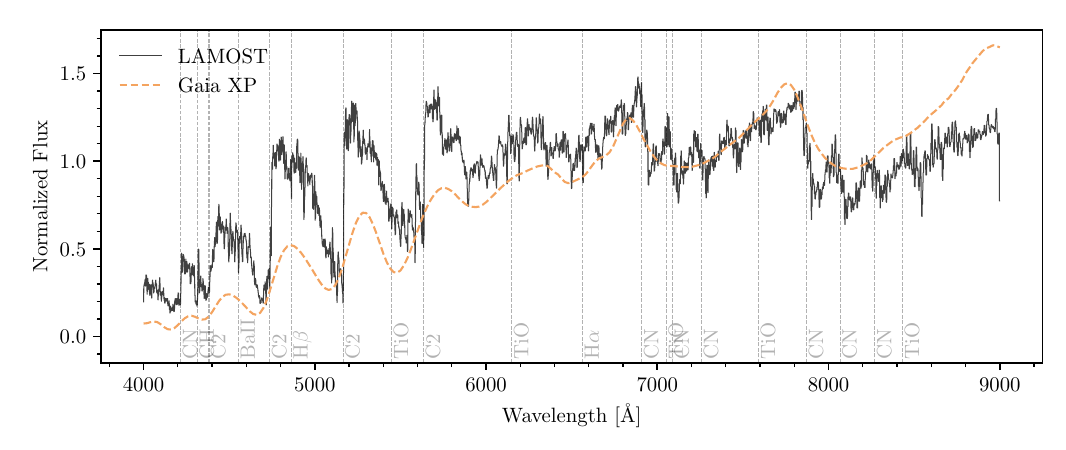}
\caption{\label{fig:LAMOST_gaia}{The LAMOST spectrum (solid black line) and corresponding Gaia XP spectrum (orange dashed line) of a previously known carbon star, 0316+1006, or Gaia DR3 14547844505574400 \citep{Totten1998,GaiaDR3}.} The wavelengths of major late-type stellar spectral features like the bandheads of CH, C$_2$, CN and TiO are marked by vertical lines.  We used plots like these to visually vet our C star training sample.}
\end{figure*}

From this sample, we removed stars with Galactic latitude $|b| < 10$ to avoid significant reddening (by dust in the Galactic plane) and overcrowded fields. We also removed stars with $G > 16.5$ to avoid spikes and other non-physical issues at low S/N in the Gaia XP spectra. In addition, we reject $\bprp < 0.7$ to exclude objects outside the temperature regime of C stars \citep{Green2019}, especially DQ white dwarfs with carbon bands. These cuts resulted in 1746 unique objects.  Finally, we visually inspected the LAMOST spectrum for each of these stars, selecting objects with significant C$_{\rm 2}$ bands (usually including visible C$_{\rm 2}$ at 5636\AA), resulting in our final pure sample of 926 carbon stars. An examples of the plots used for this visual inspection are shown in Figure\,\ref{fig:LAMOST_gaia}.

\subsection{Control Sample \label{sec:extendedCtl}}

In order for our trained machine learning algorithm to correctly classify carbon stars, it must be trained to identify objects labeled as ``not carbon star''. This motivated the need for a well defined control sample of non-carbon star objects. This control sample should contain a representative sample of objects that may found in the general Gaia data set, including objects such as O-rich stars, quasars, galaxies, and other stellar types as rare as C stars (e.g., S-type, Wolf-Rayet stars, brown dwarfs, cataclysmic variables).

We aimed to have a large control sample that would contain objects across these representative classes, but without placing many constraints that might cause unknown selection effects in the control sample that differ from the main carbon star sample. We simply selected from Gaia the first one million objects using the Gaia provided random index, and applied the same data cuts as from the carbon star sample. These cuts included magnitude, color, Galactic latitude, and that there was XP spectra available. The resulting control sample comprises 18909 random stars with the same magnitude and color range as the C star sample.

\section{Features Used in Training\label{sec:features}}

While a carefully selected and curated training sample is crucial for accurate machine learning classifications, it is equally important to choose training features that enable the best discrimination between classes of objects. Since carbon stars are anticipated to span a wide range of luminosities, from luminous AGB stars to low-mass dwarfs, and may be at any distance within the detectable range of Gaia, we have ensured that our training features are distance-independent. 

The features we selected to train on all incorporate some form of information about the important prominent spectral features found in carbon stars, as well as non-carbon stars. While these features may have overlap in the information about specific spectral features, multiple representations of the same data input to machine learning models often lead to more accurate outcomes (e.g., \citealt{Dietterich2000}).

The 133 features we have selected cover three main categories: (1) Gaia photometry in the from of the Gaia colors \citep[\bprp, $G_{BP} - G$, $G - G_{RP}$;][]{GaiaDR3}, (2) the spectral BP/RP coefficients for the Gaia XP spectra as given by Gaia, and (3) spectral indices as measured from the sampled Gaia XP spectra \citep{DeAngeli2023} . 

We include all BP and RP coefficients, for a total of 110 coefficient features. Since these coefficients encode information about the apparent magnitude, we normalized both BP and RP each to the value of the respective first coefficient. This makes the first XP coefficient in each set (BP1, RP1) both 1, and all other coefficients scaled appropriately.

We also include 20 spectral index features, which are measured from the sampled XP spectra, which are made from the XP coefficients using \texttt{GaiaXPy}. We use the standard Gaia XP sampling wavelength grid, but interpolate it onto a 1\AA\ wavelength spacing to allow the index to be measured when found between the standard sampling grid spacing. For the training data, we performed a standard sample scaling to the spectral indices individually to remove the mean and to a variance of one.

\begin{deluxetable}{lcccc}
\tablecaption{Spectral Indices \label{tab:sis}}
\tablehead{
\colhead{Spectral Index} &  \multicolumn{2}{c}{In-Band} &  \multicolumn{2}{c}{Out-of-Band} \\
\colhead{Name} & \colhead{$\lambda^{in}_1$} & \colhead{$\lambda^{in}_2$} & \colhead{$\lambda^{out}_1$}  & \colhead{$\lambda^{out}_2$} \\
& \multicolumn{4}{c}{\AA} }
\startdata
$^*$CaII\,K & 3924.8 & 3944.8 & 3944.8 & 3954.8 \\
$^*$CaII\,H & 3960.0 & 3980.0 & 3980 & 4030 \\
CaI\,4217 & 4217.9 & 4237.9 & 4237.9 & 4257.2 \\
$^*$CH & 4286.2 & 4316.2 & 4261.2 & 4286.2 \\
CN\,4216 & 4144 & 4178 & 4246 & 4284 \\
C2\,4382 & 4350.0 & 4380.0 & 4450.0 & 4600.0 \\
C2\,4737 & 4650.0 & 4730.0 & 4750.0 & 4850.0 \\
H$\beta$ & 4823.0 & 4900.0 & 4945.0 & 4980.0 \\
$^*$C2\,5165 & 5028.0 & 5165.0 & 5210.0 & 5380.0 \\
$^*$Mg I & 5154.1 & 5194.1 & 5101.4 & 5151.4 \\
TiO\,5448 & 5448 & 5500 & 5380 & 5440 \\
C2\,5636 & 5400.0 & 5630.0 & 5650.0 & 5800.0 \\
TiO\,6151 & 6151.0 & 6300.0 & 6400 & 6520 \\
H$\alpha$ & 6519.0 & 6609.0 & 6645.0 & 6700.0 \\
CN\,6926 & 6935.0 & 7035.0 & 6850.0 & 6900.0 \\
CN\,7088 & 7075.0 & 7233.0 & 7039.0 & 7075.0 \\
TiO\,7053 & 7053.0 & 7250.0 & 7340 & 7500 \\
CN\,7259 & 7233.0 & 7382.0 & 7382.0 & 7425.0 \\
TiO\,7592 & 7589.0 & 7885.0 & 7390 & 7560 \\
CN\,7872 & 7850.0 & 8050.0 & 7650.0 & 7820.0 \\
CN\,8067 & 8059.0 & 8234.0 & 8234.0 & 8263.0 \\
CN\,8270 & 8263.0 & 8423.0 & 8423.0 & 8481.0 \\
TiO\,8432 & 8432 & 8690 & 8710 & 8830 \\
TiO\,8860 & 8860 & 9000 & 9000 & 9050 \\
\enddata
\tablecomments{The designated name of each spectral index is listed,
along with the start and end wavelengths used to calculate the average
flux in the numerator (in-band) and the denominator (out-of-band) for that
index, as in Equation\,1.  Wavelengths listed with decimal places are taken from \cite{Roulston2020} for that index.
Those with no decimal place we defined by visual inspection of SDSS spectra of M or C stars.
Spectral indices marked with an asterisk were calculated, but not used in the candidate selection.}
\end{deluxetable}

We calculate a spectral index as simply a ratio that describes the relative strength of various absorption bands. This ratio is of the mean flux in-band of some spectral feature of interest, to the mean flux out-of-band in a nearby region that acts as local pseudo-continuum. Table\,\ref{tab:sis} lists the spectral indices that we consider in this work. Not all the indices listed in the table are used, with those calculated for later exploration but not used to train marked with an asterisk. For each index, we give wavelength range edges for the in-band and out-of-band flux regions we use. 

For the spectral indices listed in Table~\ref{tab:sis}, the mean of the in-band flux is the numerator, and the mean of the nearby out-of-band flux or the pseudo-continuum flux is the denominator. This is shown by the following equation,

\begin{equation}
    \textrm{Spectral Index} = \frac{\overline{ \textrm{F}_{\textrm{in-band} } } }{\overline{ \textrm{F}_{\textrm{out-of-band} } } } 
    = \frac{\int_{\lambda_1^{\text{in}}}^{\lambda_2^{\text{in}}} f(\lambda) \, d\lambda}{\int_{\lambda_1^{\text{out}}}^{\lambda_2^{\text{out}}} f(\lambda) \, d\lambda}
\end{equation}
where the wavelength limits are listed for each index in Table~\ref{tab:sis}.

The spectral indices in Table\,\ref{tab:sis} were chosen to characterize the strengths of absorption features found commonly in stellar spectra, and most particularly the C$_2$ and CN molecular bands. Many of the spectral indices were adopted directly from the PyHammer2 code of \citealt{Roulston2020}.  The strength of certain spectral indices change depending on the stellar effective temperature, gravity and atmospheric abundances. For cooler carbon AGB giants (C-N stars) with depressed blue flux, the CN bands will be the strongest defining features, whereas for warmer carbon stars (e.g., C-H and C-R types) the C$_2$ bands will be the strongest defining features.  We excluded some important molecular bands seen in C stars which are not distinguishable from other common absorption features at the Gaia XP spectral resolution.  For example, the CH band occurs frequently in normal O-rich stars from late-F through early K- types \citep{Gray2022}.  The C$_2$ $\lambda$5165\AA\ band overlaps strongly with Mg\,I and MgH absorption at $\lambda$5175\AA.  The CaII H\&K lines are evident in both O- and C-rich stars.  These features were not used as features in our classifiers.

\section{Machine Learning Optimization \label{sec:ML}}

\subsection{XGBoost  \label{sec:XG} }
 We chose the XGBoost algorithm to help identify new C stars because of its proven successful implementation with large spectroscopic surveys. Studies that utilize XGBoost for stellar classification purposes using spectroscopy and photometry include: metal-poor carbon-enhanced stars with Gaia DR3 \citep{Lucey2023}; metal-poor candidates with Gaia DR3 \citep{Yao2023}; stars, quasars, and galaxies with BASS DR3 \citep{Li2021}; red clump stars with LAMOST DR7 \citep{He2023}; symbiotic stars with LAMOST DR9 \citep{Jia2023}; M giants with LAMOST DR9 \citep{Yi2019}; M subdwarfs in LAMOST and SDSS DR7 \citep{Yue2021}, to name a few.

XGBoost is a gradient-boosting decision tree algorithm \citep{Chen2016}. This algorithm builds repeated decision trees to fit the residuals from a previous tree, until the residuals stop decreasing or the model reaches the maximum number of trees. Then, it sums the results from each tree, which are weighted by a learning rate ($\eta$), and this value is then plugged into the Sigmoid function, $\sigma(x) = 1 / (1 + e^{-x})$, to calculate the probability of the object belonging to a certain class.

To optimize the XGBoost hyperparameters, we used both the \texttt{ShuffleSplitCV} and \texttt{RandomSearchCV} from Scikit-Learn \citep{Scikit-learn} to create training and testing sets. \texttt{ShuffleSplitCV} is a random permutation cross-validation model that splits data into training and testing sets  based on the number of splits specified, testing size percentage, etc. \texttt{RandomizedSearchCV} seeks the best parameter combinations for the chosen model as it trains the model iteratively with cross validation (using \texttt{ShuffleSplitCV}) and then calculates the accuracy score for each set of parameters. It then uses the highest accuracy score and returns the best parameters for the model.

When training XGBoost, many hyperparameters can be adjusted, such as the learning rate ($\eta$), the maximum depth of a tree, and the minimum loss reduction required to make a further partition on a leaf node of the tree ($\gamma$). Using the recommended grid from \citep{Yao2023}, \texttt{RandomSearchCV} randomly selects values from a predefined grid in hyperparameter space, listed below, to find the optimal set of parameters:
 
\begin{enumerate}
 \setlength{\itemsep}{0pt}
  \item \texttt{n\_estimators}: from 900 to 1500 in steps of 50
  \item \texttt{max\_depth}: from 2 to 15 in steps of 1
  \item \texttt{learning\_rate}: from 0.05 to 1 in steps of 0.05
  \item \texttt{subsample}: from 0.5 to 1 in steps of 0.05
  \item \texttt{colsample\_bytree}: from 0.3 to 0.9 in steps of 0.05
  \item \texttt{min\_child\_weight}: from 1 to 20 in steps of 1
  \item \texttt{gamma}: from 0 to 0.7 in steps of 0.02
\end{enumerate}

This model, using the LAMOST C stars and the randomly-selected Gaia control sample, is the better-performing model in regards to purity estimates discussed in more detail in Sec. ~\ref{sec:FAST}.

\subsection{Random Forest   \label{sec:RF}}

The Random Forest algorithm (RF) has been commonly used and evaluated for stellar classification in astronomy \citep{Solorio-ramirez2023}. Studies that use multi-dimensional data such as spectroscopic and photometric observations with RF for the purpose of stellar classification include: WDs with Gaia DR3 \citep{Garcia-Zamora_2024}; M dwarfs with SDSS and 2MASS \citep{Sithajan2023}; RR Lyrae stars with SDSS DR15 and Gaia DR2 \citep{Zhang2020}; to name a few.

The architecture of the Random Forest begins with a single decision tree created from a subset of the training data. Each data point goes through a series of nodes, or decision points, where it will branch off to the left or right depending on the criteria of the node.
At each node in a decision tree, a subset of features is randomly selected. The algorithm then decides which feature from this subset will be used to split the data.  This selection aims to find the feature that best separates the data into different classes, and to
 decide on the optimal threshold value for the splitting  that results in the greatest possible separation between classes.
Each path will lead to a series of other nodes that will be iterated across until a prediction about the data can be eventually reached. The algorithm itself contains numerous trees that are each made with a different subset of data, so they each have different predictions. The Random Forest is then able to give a likelihood score for each source given as a percentage of the trees that came to the same conclusion.

We used the Random Forest algorithm from the Scikit-Learn library. To create a random cross-validation subset, we used the \texttt{train\_test\_split} function. We chose a training set size of .75, meaning 25\% of the original training data is randomly chosen to be set aside to test the accuracy of the model rather than being used to create the algorithm. 

Many hyperparameters were tested to improve the accuracy of the RF model; the two hyperparameters with noticeable positive effects on the model's metrics were \texttt{max\_features} and \texttt{bootstrap}. The \texttt{max\_features} hyperparameter tells the algorithm how many features it can use to make a decision at each node. We found that the optimal value of \texttt{max\_features} is 19. The \texttt{bootstrap} hyperparameter is what tells the algorithm to use only a fraction of the objects for each decision tree. By setting this parameter to false and letting each decision tree be created with all stars, the model is more accurate even if it is more computationally expensive. The remaining hyperparameters were left at default values.

\subsection{Feature Importance}

\begin{figure*}
\centering
\includegraphics[width=\textwidth]{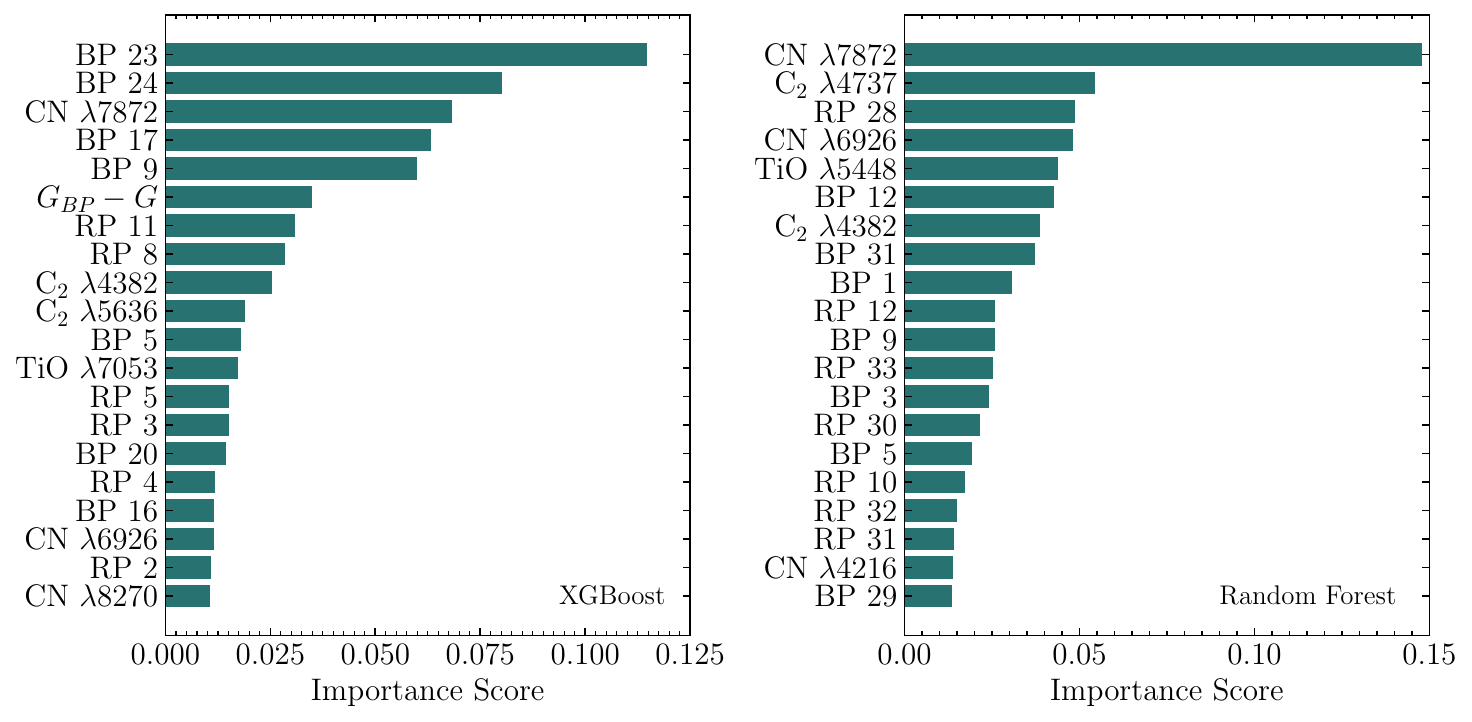}
\caption{\label{fig:feature_importance}{Importance scores of features used in XGBoost (left) and Random Forest (right), for the 20 most important features in each.  Features marked (BP $n$) or (RP $n$) are the $n$th Hermite polynomial coefficients of the BP/RP halves of the XP spectra.  Spectral indicies are shown with the central wavelength for that feature, which are listed in Table\,\ref{tab:sis}.  
} }
\end{figure*}

The relative importance of features in both the XGBoost and RF training are shown in Fig \,\ref{fig:feature_importance}. In XGBoost, feature importance is measured by `gain' which quantifies the average improvement in the loss function achieved by splits using a given feature. In RF, Gini importance is used which computes the total reduction in Gini impurity across all trees for splits involving a specific feature. These two methods of deriving feature importance, while not the same, adequately rank the effectiveness of a specific feature in the classification of an object within the model. 

The two algorithms prefer remarkably different features for selecting C stars, but they do share the SIs for CN$\lambda$7872, C$_2\lambda$4382 and CN$\lambda$6926, as well as the polynomial coefficients BP 5,  and BP 9. The most important features used in XGBoost include several BP spectral coefficients, CN\,$\lambda$7872 and C$_2\,\lambda$4382, and the color $G_{BP}-G$. The most important feature for RF by far is the spectral index CN\,$\lambda$7872. The other SI features with high importance scores are C$_2\lambda$4737, CN$\lambda$6926, TiO$\lambda5448$ , and C$_2 \lambda$4382. RF does not yield high importance scores for any colors. Identifying SIs associated with carbon bands as key features in these models validates our choice of specified SIs for classifying C stars.

\section{Selection Results  \label{sec:SelectionResults}}

With our optimized machine learning models, we trained both the XGBoost and RF models with the same data comprised of our C star sample and the control sample. With the trained models, we proceeded to predict classifications (C, or non-C) on all possible objects that meet the same cuts we placed on the training data. Of the 219,197,643 objects in Gaia DR3 with XP spectra, we predicted labels for the 99,575,842 objects that meet our cuts of $G<16.5$ and  $\bprp > 0.7$.

The XGBoost model labeled 131,265 objects as a C star, and the RF model labeled 130,437 objects as a C star. While each of these sets of predictions likely have their own set of biases, purities, and completeness, we aim to select a large sample with the highest purity to aid in our calculations of the dC space density. We therefore cross-matched these two sets of predictions, and select only objects that appear in both the XGBoost and RF predictions as a C star. This results in our final C star candidate sample of 43,574 objects. This is a C star candidate fraction of $0.04\%$ of the tested sample.

We reemphasize that our selection of C star candidates is based only on spectral shape and spectral features, irrespective of either apparent or absolute magnitude.  Because of that, the vast majority of C stars in our catalog are giants at relatively large distances, including about 10,800 in the LMC and 3,000 in the SMC.  Among candidates in the Milky Way with reliable parallaxes, the catalog contains about another 10,000 giants, and about a thousand dwarf C stars.

The full data tables we used for the training, control and candidate samples (including SIs and BP/RP coefficients) are available on request from the authors.
\movetabledown=75mm
\begin{rotatetable*}
\begin{deluxetable*}{lrrrrrrrrrrrrrrrrrrrrrrrlr}
\tabletypesize{\scriptsize}
\tablecaption{C Star Candidates \label{tab:candidates}}
\tablehead{
\colhead{Source ID} & \colhead{R.A.} & \colhead{Dec} &
\colhead{$\pi$} & \colhead{$\sigma_\pi$} & \colhead{$G$} &
\colhead{$B_P$} & \colhead{$R_P$} & \colhead{P$_{XG}$} &
\colhead{P$_{RF}$} &
\colhead{$r_{\text{med\_geo}}$} & \colhead{$r_{\text{lo\_geo}}$} & \colhead{$r_{\text{hi\_geo}}$} & 
\colhead{$A_V$} & \colhead{$E(B-V)$} & \colhead{$G_0$} & 
\colhead{$B_{p,0}$} & \colhead{$R_{p,0}$} & \colhead{$M_{G,0}$} & 
\colhead{X} & \colhead{Y} & \colhead{Z} & 
\colhead{FAST Date} & \colhead{VI Score} & \colhead{Spectral Type} & 
\colhead{Reference} \\
\colhead{} & \colhead{} & \colhead{} & \colhead{(mas)} & \colhead{(mas)} & \multicolumn{3}{c}{(mag)} & \colhead{} & \colhead{} & \multicolumn{3}{c}{(pc)} & \multicolumn{6}{c}{(mag)} & \multicolumn{3}{c}{(pc)} & \colhead{} & \colhead{} & \colhead{} & \colhead{}
}
\startdata
1272448486574848 & 44.15446 & 1.94882 & 0.657 & 0.052 & 16.397 & 16.839 & 15.800 & 0.826 & 0.58 & 1469 & 1373 & 1601 & 0.247 & 0.080 & 16.191 & 16.572 & 15.644 & 5.348 & -9103 & 103 & -1074 & & 9 & & \\
4914061062362752 & 41.14519 & 3.87731 & -0.003 & 0.060 & 16.484 & 16.943 & 15.860 & 0.561 & 0.57 & 7051 & 5522 & 8930 & 0.027 & 0.009 & 16.461 & 16.914 & 15.843 & 2.706 & -11781 & 731 & -4203 & & 9 & & \\
6161152061519488 & 40.87365 & 5.54756 & 0.539 & 0.059 & 16.468 & 16.871 & 15.900 & 0.584 & 0.51 & 1777 & 1605 & 1939 & 0.137 & 0.044 & 16.353 & 16.722 & 15.813 & 5.191 & -9247 & 263 & -1236 & & 9 & & \\
6571063740310784 & 41.10724 & 6.02771 & 0.291 & 0.014 & 13.219 & 13.829 & 12.479 & 0.994 & 0.55 & 2999 & 2904 & 3119 & 0.329 & 0.106 & 12.944 & 13.473 & 12.270 & 0.575 & -10101 & 469 & -2152 & & 9 & & \\
9216969753176832 & 51.70675 & 6.20793 & 0.047 & 0.050 & 16.303 & 17.250 & 15.361 & 0.994 & 0.71 & 7007 & 5598 & 9154 & 0.630 & 0.203 & 15.776 & 16.567 & 14.961 & 1.574 & -13449 & 263 & -4396 & 2024.0109 & 5 & & \\
14547844505574400 & 49.71319 & 10.28331 & 0.208 & 0.015 & 13.009 & 14.093 & 11.995 & 0.999 & 0.62 & 4129 & 3924 & 4354 & 1.151 & 0.371 & 12.047 & 12.846 & 11.264 & -1.033 & -11337 & 470 & -2528 & & 9 & C-H & 1998MNRAS.294....1T \\
17073392650316800 & 50.47997 & 11.91309 & -0.187 & 0.049 & 15.919 & 17.170 & 14.841 & 0.999 & 0.66 & 14166 & 11513 & 17612 & 1.151 & 0.371 & 14.956 & 15.923 & 14.110 & -0.838 & -19582 & 1818 & -8537 & & 9 & & \\
17882289611115264 & 50.60539 & 13.52878 & 0.040 & 0.059 & 15.876 & 17.080 & 14.808 & 0.985 & 0.56 & 6049 & 4675 & 7866 & 0.877 & 0.283 & 15.143 & 16.130 & 14.252 & 1.301 & -12843 & 853 & -3355 & & 9 & & \\
22312526901938432 & 41.02680 & 10.64374 & 0.064 & 0.058 & 16.261 & 16.851 & 15.536 & 1.000 & 0.55 & 5894 & 4769 & 7771 & 0.466 & 0.150 & 15.872 & 16.346 & 15.241 & 1.936 & -12381 & 1331 & -4175 & & 9 & & \\
23995299383594880 & 34.80424 & 9.857804 & 0.194 & 0.018 & 12.520 & 13.194 & 11.744 & 1.000 & 0.68 & 4734 & 4362 & 5155 & 0.000 & 0.000 & 12.520 & 13.194 & 11.744 & -0.707 & -10863 & 1226 & -3221 & & 9 & & \\
25063921606595200 & 38.93471 & 10.51698 & -0.032 & 0.054 & 16.364 & 17.040 & 15.585 & 0.998 & 0.55 & 8931 & 6966 & 12329 & 0.356 & 0.115 & 16.066 & 16.654 & 15.359 & 2.054 & -12384 & 1519 & -4427 & & 9 & & \\
28989727873584256 & 45.70616 & 13.72115 & 0.023 & 0.026 & 13.999 & 14.700 & 13.203 & 1.000 & 0.69 & 10924 & 8720 & 14737 & 0.356 & 0.115 & 13.701 & 14.314 & 12.977 & -0.818 & -14219 & 1627 & -4917 & & 9 & C-H & 2016ApJS..226....1J \\
30858038647425664 & 46.74438 & 14.87337 & 0.305 & 0.019 & 13.242 & 13.914 & 12.459 & 1.000 & 0.52 & 2872 & 2731 & 3002 & 0.493 & 0.159 & 12.830 & 13.380 & 12.145 & 0.529 & -10363 & 594 & -1696 & & 9 & & \\
34334969292335488 & 45.04717 & 16.82782 & 0.022 & 0.032 & 14.754 & 16.058 & 13.694 & 0.970 & 0.75 & 9283 & 7667 & 11583 & 0.850 & 0.274 & 14.044 & 15.138 & 13.154 & -0.861 & -15492 & 2383 & -5595 & & 9 & C & 2007A\&A...475..843M \\
36741456649738880 & 58.56773 & 11.60577 & -0.033 & 0.024 & 14.756 & 16.118 & 13.619 & 0.987 & 0.63 & 15663 & 12221 & 20589 & 0.932 & 0.301 & 13.977 & 15.109 & 13.027 & -2.203 & -22878 & 532 & -8847 & & 9 & C-N & 2016ApJS..226....1J \\
38080764891488768 & 55.43738 & 13.20372 & 0.025 & 0.027 & 14.647 & 15.484 & 13.753 & 0.987 & 0.71 & 9486 & 7681 & 12041 & 0.850 & 0.274 & 13.936 & 14.563 & 13.213 & -0.198 & -13777 & 583 & -3548 & & 9 & & \\
42005467350070656 & 51.74965 & 14.66583 & 0.117 & 0.024 & 13.890 & 16.135 & 12.580 & 0.991 & 0.69 & 5334 & 4659 & 6142 & 0.767 & 0.248 & 13.248 & 15.304 & 12.093 & -0.457 & -12640 & 810 & -3023 & & 9 & C & 2000PASP..112.1315L \\
42217299432439168 & 51.94922 & 15.53781 & 0.990 & 0.048 & 16.257 & 16.846 & 15.543 & 0.566 & 0.62 & 980 & 934 & 1032 & 0.548 & 0.177 & 15.798 & 16.252 & 15.195 & 5.867 & -8922 & 151 & -503 & & 9 & & \\
45546070886222592 & 61.70354 & 16.47176 & 0.131 & 0.022 & 12.764 & 14.273 & 11.619 & 0.999 & 0.66 & 6143 & 5495 & 7157 & 1.069 & 0.345 & 11.870 & 13.115 & 10.940 & -1.965 & -13389 & 355 & -2494 & & 9 & C & 2007A\&A...475..843M \\
48568216034333696 & 64.12551 & 19.13301 & 0.090 & 0.024 & 14.500 & 15.509 & 13.517 & 0.987 & 0.57 & 6699 & 5802 & 8021 & 1.261 & 0.407 & 13.446 & 14.144 & 12.716 & -0.710 & -14390 & 477 & -2516 & & 9 & C-R & 2016ApJS..226....1J \\
49546850103156480 & 59.10374 & 17.18282 & 0.030 & 0.041 & 15.750 & 16.626 & 14.833 & 0.931 & 0.60 & 8491 & 6242 & 10869 & 0.959 & 0.309 & 14.948 & 15.587 & 14.224 & -0.398 & -18526 & 1154 & -5269 & & 9 & & \\
50010947089387520 & 59.56182 & 18.83563 & 0.229 & 0.025 & 14.149 & 14.931 & 13.291 & 1.000 & 0.56 & 3702 & 3401 & 4246 & 0.822 & 0.265 & 13.461 & 14.040 & 12.769 & 0.702 & -11315 & 410 & -1504 & 2023.1109 & 2 & & \\
51532018347626496 & 58.09941 & 20.37700 & 0.022 & 0.021 & 14.412 & 15.308 & 13.490 & 0.999 & 0.70 & 10757 & 9237 & 13270 & 0.630 & 0.203 & 13.885 & 14.625 & 13.090 & -1.162 & -17238 & 1542 & -4337 & & 9 & & \\
51591907371533568 & 58.29854 & 20.72630 & 0.014 & 0.049 & 15.940 & 16.684 & 15.115 & 0.995 & 0.66 & 7702 & 5973 & 10212 & 0.548 & 0.177 & 15.482 & 16.090 & 14.767 & 0.953 & -15318 & 1233 & -3362 & & 9 & & \\
52881325270649472 & 62.87598 & 22.91861 & 0.847 & 0.032 & 11.300 & 12.001 & 10.464 & 0.963 & 0.69 & 1146 & 1105 & 1187 & 0.795 & 0.256 & 10.635 & 11.140 & 9.959 & 0.357 & -9176 & 151 & -373 & & 9 & & \\
53795466114074368 & 61.22232 & 23.23628 & 2.525 & 0.034 & 10.463 & 11.049 & 9.738 & 1.000 & 0.59 & 391 & 386 & 396 & 0.411 & 0.130 & 10.052 & 10.562 & 9.473 & 0.026 & -6158 & 120 & -215 & & 9 & & \\
\enddata
\tablecomments{Table 1 is published in its entirety in the
  machine-readable format. A portion is shown here for guidance
  regarding its form and content. R.A. and Dec are given in decimal degrees (J2000 epoch).}
\end{deluxetable*}
\end{rotatetable*}

While we investigate our selection purity and completeness in the next sections, our initial cursory check was to cross-match our selected C star candidates to available large spectroscopic surveys. 

Cross-matching the candidate list to the approximately 10 million LAMOST DR9 spectra yields 1646 matches. Of those matches, 1,215 (74\%) are classified as LAMOST C stars, 144 (8.7\%) as G stars, 77 (4.7\%) as K stars, with the remainder (13\%) scattered across other stellar types. Some of the stars that LAMOST classifies as G and K type may have weak carbon features, which could indicate a purity as high as 87\%.  Note that of these 1,215 LAMOST DR9 C stars our algorithms selected, 545 were included in our LAMOST DR8v2.0 training sample.  

Cross matching our same candidate sample to approximately 3.2 million SDSS spectra yields 380 matches. Of those, 298 (78.4\%) are classified by SDSS as C stars, 11 (2.9\%) as G stars, and 49 (12.9\%) as K stars. Again, if the G and K have weak carbon features, our purity could be up to 94\%. 

\begin{figure}
\epsscale{1.2}
\plotone{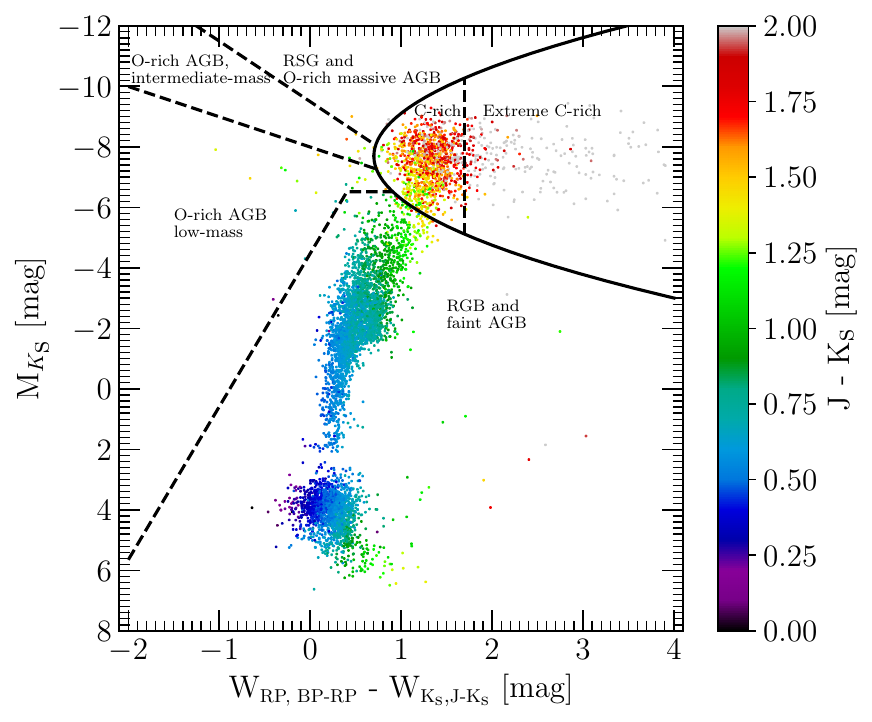}
\caption{Revised Gaia-2MASS plot from \citet{Lebzelter2018} showing our C star candidate sample. We show the boundaries from \citet{Lebzelter2018} in dashed black lines, and are labeled accordingly. Comparing our plot to the updated version found in \citet{Abia2022}, we see that our C star candidates span all types, including N, J, R-hot, and dwarf carbon stars.}
\label{fig:Gaia-2MASS}
\end{figure}

Figure \ref{fig:Gaia-2MASS} presents a revised version of the main figure from \citet{Lebzelter2018}, incorporating our comprehensive sample of C star candidates. The original plot in \citet{Lebzelter2018} was intended to distinguish between AGB stars and regions enriched in C and O elements. However, a newer version of the plot in \citet{Abia2022} highlights the presence of C-rich stars in regions beyond the labeled C-rich region. By comparing our sample to the plot from \citet{Abia2022}, we infer that our sample encompasses a diverse range of C star types, including N, J, R-hot, and dwarf carbon stars.

\subsection{Completeness  \label{sec:completeness}}

\begin{figure*}
    \centering
\includegraphics[width=0.497\textwidth]{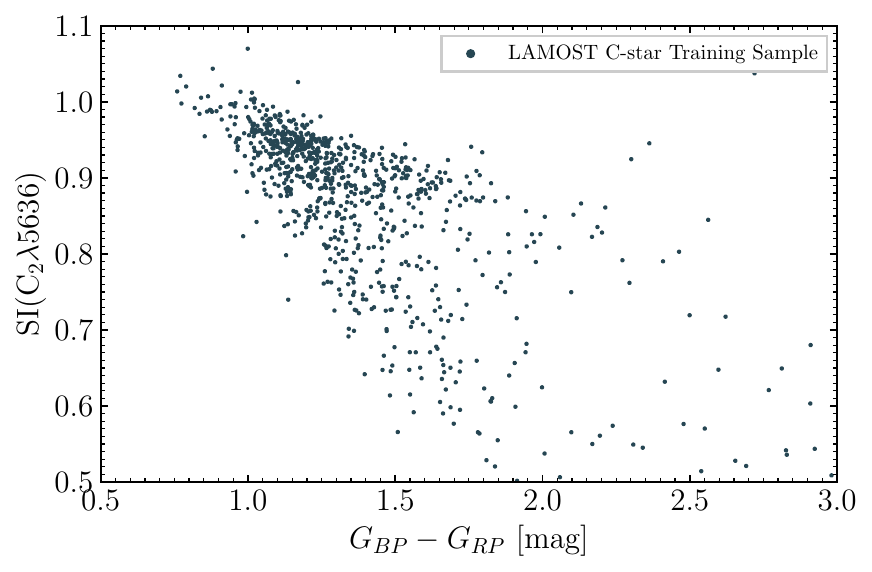}
\includegraphics[width=0.497\textwidth]{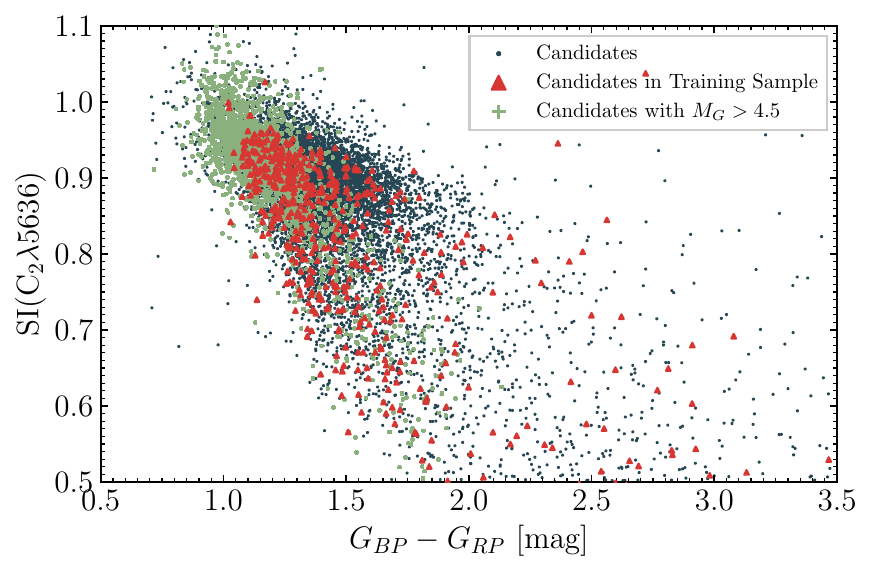}
\includegraphics[width=0.5\textwidth]{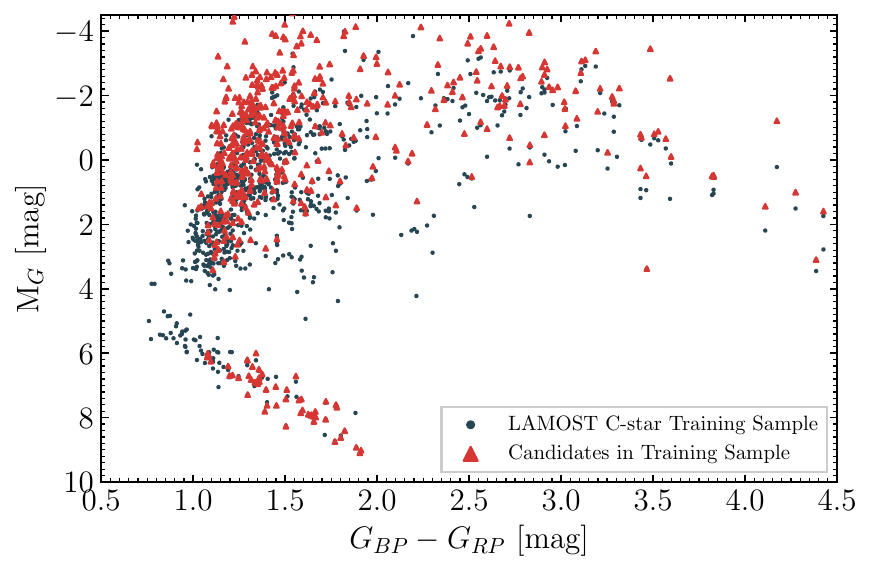}
\caption{\label{fig:TrainingSampleRecovery}{Comparison of our candidate sample to our LAMOST C star training sample.  UPPER LEFT:  C$_2\,\lambda 5636$ spectral index measurements vs. \bprp\, color for our full training sample of 926 visually vetted C stars from LAMOST. UPPER RIGHT: Same plot for objects selected by our method as C star candidates. Blue points are all 11,527 candidates in uncrowded sky regions (excluding the Galactic plane and the LMC/SMC regions). Red points show the 592 (63.9\% of all) the LAMOST training sample selected as candidates. Green points show all candidates with $M_G>4.5$ in uncrowded regions.  The lowest completeness is in the region of bluer C stars with weaker C$_2$ bands, which are predominantly dCs. BOTTOM: Color-magnitude diagram for the full LAMOST training sample (blue points), showing those selected as C star candidates in red.  Again, the bluest C stars show greater incompleteness, with few LAMOST dCs selected between $4.5<M_G<6$.
}}
\end{figure*}

The ideal completeness check would be to match our candidate C star list against a large sample of C stars with reliable spectroscopic classifications at higher resolution, which also have Gaia XP spectra in the same magnitude range as our survey.
In fact, the best such comparison sample is our LAMOST C star training sample (see Section\,\ref{sec:cstarTraining}).  This is clearly not ideal, because our methods should detect most of those by definition. Therefore, a completeness test on this sample probably allows an {\em upper limit} to our actual completeness.  Figure\,\ref{fig:TrainingSampleRecovery} shows the properties of the full training sample (926 LAMOST C stars vetted by eye with scores of 4 or 5; see Section\,\ref{sec:vi}).  Just 592 (63.9\% of) the LAMOST training sample are selected as candidates.  The incompleteness is worst in the region of bluer C stars with weaker C$_2$ bands, which are predominantly dCs.

We certainly should expect greater completeness for our own training sample. It's possible that we do not not retrieve the full sample because of model overfitting.  For the blue dCs, carbon features are weak, and easily confused with features commonly seen in normal mid-G to early K-type stars.  For such early-type dCs, the model may be focusing on irrelevant features or noise. Such stars may also be near the decision boundary, and therefore not meet the retrieval criteria.

Table\,\ref{tab:completeness} shows the percentage of LAMOST stars retrieved as candidates, as a function of absolute $G$ band magnitude $M_G$, in thirteen 1\,mag bins from --3.5 to 9.5.  Essentially none of the bluest, most luminous dCs are selected as candidates in the range 4.5$<M_G<$5.5.  By contrast, completeness is excellent for red C stars toward both the faint and luminous ends of the C star sequence. For the full range of $M_G$, completeness is 63.8\%, similar to e.g., a recent Gaia search for CEMP stars \citep{Lucey2023}.

\begin{deluxetable}{cccccc}
\tablecaption{Completeness by $M_G$\label{tab:completeness}}
\tablewidth{0pt}
\tablehead{
\colhead{\( M_{G_1} \)} & \colhead{\( M_{G_2} \)} & \colhead{\( N_{\text{LAMOST}} \)} & \colhead{\( N_{\text{Cand}} \)} & \colhead{\%}
}
\startdata
-3.5 & -2.5 &  20 & 20  & 100.0 \\
-2.5 & -1.5 &  77 & 77  & 100.0 \\
-1.5 & -0.5 & 123 & 118 &  95.9 \\
-0.5 &  0.5 & 164 & 127 &  77.4 \\
 0.5 &  1.5 & 187 & 117 &  62.5 \\
 1.5 &  2.5 & 137 &  58 &  42.3 \\
 2.5 &  3.5 &  93 &  20 &  21.5 \\
 3.5 &  4.5 &  14 &   1 &   7.1 \\
 4.5 &  5.5 &  18 &   0 &   0.0 \\
 5.5 &  6.5 &  38 &   9 &  23.7 \\
 6.5 &  7.5 &  27 &  19 &  70.4 \\
 7.5 &  8.5 &  19 &  18 &  94.7 \\
 8.5 &  9.5 &   6 &   5 &  83.3 \\
\enddata
\tablecomments{Columns are (1) Bright edge of the absolute $G$ magnitude bin (2) faint edge (3) the number of LAMOST C stars in our training sample (4) the number of such stars retrieved as candidates by our selection algorithm (5) the completeness i.e., the percentage of LAMOST C stars retrieved as C star candidates.  }
\end{deluxetable}

For definitive C giants, with $M_G\leq2.5$, 73\% (517) of the 708 LAMOST C stars are retrieved. For definitive dCs, with $M_G>4.5$, 47.2\% (51) of the 108 dCs are retrieved. Higher completeness for dCs is achieved only by excluding the blue, luminous objects e.g., 56.7\% for $M_G>5.5$ or 80.7\% for $M_G>6.5$.

To test the completeness of our selection specifically for giant C stars in the literature, we cross-correlate our list of candidates against Table 1 of \citep{Abia2022}, a list of Galactic carbon or related stars culled from the literature, including C-N and C-J type AGB stars, as well as warm C-R stars.  Our candidate list includes 92\% of the cooler C stars in this list (374 stars with N and J types).  As expected, we retrieve a much smaller fraction (35\%) of the (234) warm C-R stars in their list; such stars have weak C$_2$ and CN bands (see e.g., Fig 1 of \citealt{Ji2016}).

\section{Spectroscopic Verification of Candidates \label{sec:FAST}}

To verify whether our candidates are actually C stars, and thereby estimate the purity of this new sample, we selected a random subset of candidates for follow-up spectroscopic observations at higher spectral resolution to confirm the existence of carbon molecular band features. We obtained spectra of a subset of the candidate list over 5 observing runs at the Fred Lawrence Whipple Observatory (FLWO) on Mt. Hopkins in southern Arizona.  The spectra were obtained using the FLWO 1.5m Tillinghast telescope with the FAST spectrograph. For the first four runs, spectra were taken using a 3\arcsec\, slit.  For the last run, we used a 1.5\arcsec\ slit.  A tilt of 625 for the 300 lines$/$mm grating yielded a wavelength range of about 3850 to 7850 \AA, with a typical FWHM resolution of 6–7 \AA. The total exposure time for a target spectrum depended on the object's apparent magnitude, ranging from 10 to 400 seconds.  The observing dates and number of C star candidates observed are listed in Table\,\ref{tab:fastruns}

Target selection differed across runs. For all FAST runs, we selected candidates by training the XGBoost and Random Forest algorithms, generating a candidate C star sample from their overlap.  For the FAST runs in 2023, we used our initial training samples described in the Appendix\,\ref{sec:initCtraining}, with 2315 C stars selected from the literature, and an extended control sample of 18,914 objects (from which we had removed 17 C star candidates). 

Before the Oct 2023 run, we trained on all SIs (Table\,\ref{tab:sis}) along with all the polynomial coefficients. The overlap of XGBoost and RF results yielded 212,381 candidates across the whole sky.  
Of these, we randomly selected a target list, of which 102 unique objects were observed with the FLWO\,1.5m/FAST.

While the coefficients are only available for Gaia spectra, SIs would be applicable to any spectroscopic survey, so for the Nov 2023 run, we used the same training samples, but tested training without use of coefficients.  We also dropped some of the SIs (those marked with an asterisk in Table\,\ref{tab:sis}) that showed low importance in the modeling, and were deemed likely to reduce purity:  Ca\,II, CH $\lambda$4314, C$_2$ $\lambda$5165, and Mg\,I $\lambda$5175 because at low-resolution these features overlap with features in O-rich stars.  Candidates selected both by XGBoost and RF results yielded 646,121 candidates, i.e., far less selective.  From a randomly-selected sample of targets, 274 unique objects were observed.

Before the Jan 2024 run, we used the new, final training sample, taken entirely from VI-confirmed LAMOST spectra, and trained on the reduced list of SIs, including coefficients.  The overlap sample from the XGBoost and RF algorithms included 43,574 candidate C stars.  Of these, 18,239 are in crowded regions (4,430 have $|b|\leq 10$deg and 13,808 are within 5\,deg of the Large or Small Magellanic Clouds), leaving 11,527 as our prime sample of interest for this paper. A random subsample of 333 such stars were observed with FAST in Jan 2024.

For the Mar 2024 run, we selected C star candidates from the same list as for Jan 2024.  A random subsample of 241 unique candidates were observed.

For the May 2024 run, we restricted to dC candidates only (those with $M_G>4.5$, calculated from the observed $G$ magnitudes and using the median distance \texttt{r\_med\_geo} from \citealt{Bailer-Jones2021}), with a higher 
XGProb\_C value above 0.85.  Furthermore, we sought multiple epoch observations for those dC candidates that show strong variability in ZTF lightcurves, especially if periodicity was detected.  Spectra were obtained for 125 unique dC candidates.  Given the additional selection criteria for this run, these objects must be analyzed separately for purity.  

Accounting for all 5 FAST runs, we obtained 1,147 spectra of 1,051 unique C star candidates.
Of these, 733 are included in our final candidates catalog.  Of those, 621 spectra were obtained before the 2024 May run, so are randomly selected and valid for purity calculations spanning all type of C stars below.  Of those, 565 are randomly selected in uncrowded sky regions ($|b|>10^{\circ}$ and not in the LMC or SMC).

All FAST spectra were reduced using standard IRAF\footnote{IRAF is distributed by the National Optical Astronomy Observatory, which is operated by the Association of Universities for Research in Astronomy (AURA) under cooperative agreement with the National Science Foundation.} routines \citep{Tody1986, Tody1993} as implemented in Pyraf \citep{Pyraf}. The spectra were flux calibrated using standard stars observed on the same night. 

\begin{table}
\centering
\caption{FLWO\,1.5m/FAST Observing Runs}
\begin{tabular}{l c }
\toprule
\textrm{Dates} & \textrm{Candidates Observed}\\
\hline
2023 October 14 -  16 & 102  \\
2023 November 08 - 12 & 274  \\
2024 January 08 -  15 & 333 \\
2024 March 06 - 11 & 241 \\
2024 May 06 - 10 & 125 \\
\hline
\end{tabular}
\label{tab:fastruns}
\end{table}

\subsection{Visual Inspection and Sample Purity \label{sec:vi}}

\begin{figure*}
\centering
\includegraphics[width=\textwidth]{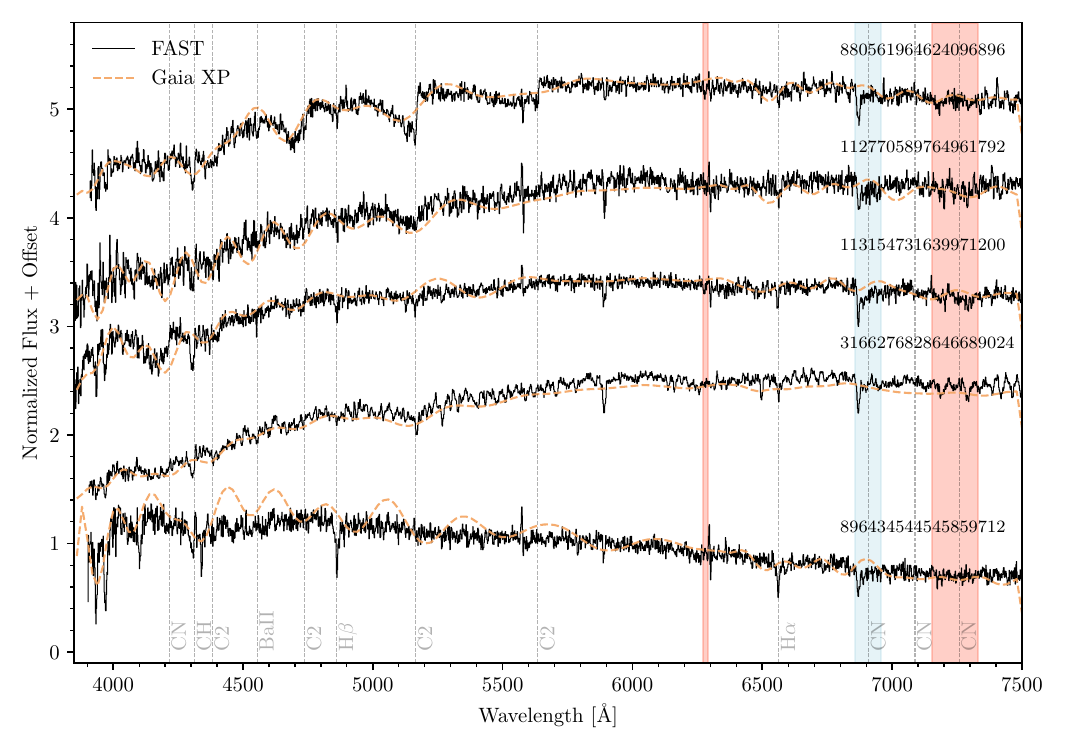}
\caption{\label{fig:FASTvi}{Example of FAST spectra with Visual Inspection scores from 5 (top) to 1 (bottom). 
The spectra are flux-calibrated using a standard star from the same night.  The Gaia spectrum for each one is overplotted in dashed orange, making it easy to see both Gaia's much lower resolution and the ringing that can occur, related to the Gaia spectral representation using Hermite polynomials. The wavelength of important bands of C$_2$ and CN, Balmer lines and Ba\,II are indicated with light vertical lines. Red and grey shaded regions correspond to strong telluric absorption (uncorrected) of H$_2$O and O$_2$. } }
\end{figure*}

We visually inspected every FAST spectrum and gave it a `VI score' from 0 to 5, based mostly on the appearance and strength of C$_2$ bands. The meaning of the scores is summarized in Table\,\ref{tab:viscoring}. An example of the plots used for visual inspection is shown in Fig\,\ref{fig:FASTvi}.
Spectra with clear, sharp C$_2$\,$\lambda$5636 features always featured other strong C$_2$ bands, and ranked the highest score of 5.  If C$_2$\,$\lambda$5636 was weak, but several other C$_2$ features were clearly visible, a spectrum ranked 4.  Spectra with weaker C$_2$ features are more common in bluer stars.  Many spectra, ranked with a VIscore of 3, show no evident C$_2\,\lambda 5636$, and weak or absent C$_2\,\lambda 4737$ but normally have clear CH$\lambda\,4315$, and evident CN$\lambda\,4216$. We generally excluded consideration of the C$_2$\,$\lambda$5165 feature, because it can look quite similar in strength and even shape to the MgH band commonly found in normal stars from mid-G to early M. 

\begin{table}
\centering
\caption{Visual Inspection Scoring \label{tab:viscoring}}
\begin{tabular}{l c r}
\toprule
\textrm{Score} & \textrm{Definition} & Number of Stars\\
               &                      & in Final Sample \\
\hline
 5 & obvious C star  & 331 \\
 4 & very likely C star  & 94 \\
 3 & probable C star  & 47 \\
 2 & possible C star  &  81 \\
 1 & probably not a C star  & 4 \\
 0 & def not a C star  & 4 \\
Total & & 565 \\
\hline
\end{tabular}
\end{table}

Of the raw final candidate sample, visual inspection of 621 FAST spectra yield 76.1\% with scores of 4 or 5, i.e., clearly C stars.  

If we include scores of 3 or higher, the purity becomes 83.8\%.  We argue that including VI scores $\geq$3  is probably reasonable, because such stars almost always include the CH band, along with at least one believable C$_2$ and/or CN bandhead. 

For analysis of what features might be used to achieve higher purity, we can include all FAST spectra of C star candidates from any of our observing runs, but the final statistics on purity should be obtained using only the randomly-selected sample of 621 candidates. Further, for most of our analysis, we will focus on the 565 that are in uncrowded, low extinction regions ($|b|>10^{\circ}$ and non-LMC/SMC).  Of those, 241 are dCs ($M_G>4.5$).  
While one might expect the machine learning algorithms to do the optimal job of selecting candidates, the training set that we provide may not represent the breadth - much less the relative prevalence - of actual C stars in the sky.  
After looking at dozens of plots of FAST VI score vs. features, we find that one of the most effective filters to enhance the purity of the sample (i.e., removing mostly those C star candidates with low VI scores) is the XGBoost C star classification probability, which we label `XGProb\_C'.  Other useful filters include features with relatively high importance based on the algorithm outputs, like spectral indices for C$_2\lambda 4737$, C$_2 \lambda 4636$, CN$\lambda 7088$, CN$\lambda 7872$, or the XP spectral coefficient BP 23.  Since we provide these features in a machine readable table, the reader can choose their own preferred filters.  We experimented with several filter combinations some of which are listed in Table\,\ref{tab:purity_filters}, to investigate their effect on the purity of the sample, based on VI scores of our FAST spectra.

\begin{table*}
\centering
\caption{Example Purity Filters \label{tab:purity_filters}}
\begin{tabular}{ccccccc}
  \hline
 & XGProb\_C  & \multicolumn{4}{c}{Spectral Index} & bp\_23\\
 &         & C$_2\,\lambda$4737  & C$_2\,\lambda$5636  & CN$\,\lambda$7088  & CN$\,\lambda$7872 &  \\
 & $>0.85$ &  $<0.6$ &  $<0.8$ &  $>1.03$ &  $<0.85$ & $<-1.5$\\
\hline
a & \checkmark & . & . & . & . & . \\
b & \checkmark & \checkmark & . & . & . & . \\
c & \checkmark & . & \checkmark & . & . & . \\
d & \checkmark & . & . & \checkmark & . & . \\
e & \checkmark & . & . & . & \checkmark & . \\
f & . & . & . & . & \checkmark & \checkmark \\
g & \checkmark & . & . & . & . & \checkmark \\
\hline
\end{tabular}
\end{table*}

In Table\,\ref{tab:VIscoresAll}, we show the number of stars in our candidate sample by VI score, for each of these 7 example filters.  We exclude candidates within crowded regions
($|b| < 10$deg or in the LMC/SMC), leaving 565 randomly-selected candidates with FAST spectra and VI scores.  Of these, 83.5\% are bona fide C stars with VI scores $\geq$3.  This is a good general number to know for the full candidate sample, irrespective of absolute magnitude or any other filters.  If we restrict the candidate sample further to have XGProb\_C$>0.85$, the purity becomes 95.5\%, albeit with the loss of 10\% of the initial candidates having VI score $\geq 3$.

In Table\,\ref{tab:dCVIscores}, we show the number of stars in our candidate dC sample by VI score, for each of the 7 example filters. All the same sky region exclusions apply as in the previous table, but in addition we impose a requirement that $M_G>4.5$.
Table\,\ref{tab:dCPurityTests} shows purity for dCs ($M_G\geq 4.5$ in uncrowded sky regions) only, based on visual inspection scores for FAST spectra in our candidate sample, with either purity filter, or with filter (a) (XGProb\_C $\geq$0.85).  We show object counts and purity fractions in 5 bins of absolute magnitude for FAST-observed candidates, counting all dCs, those with VI score$\geq 4$ or $\geq 3$. For dCs with $M_G\geq 5.5$ and VI score $\geq 3$, the sample purity is 94.8\%.  If we select all candidates in uncrowded sky regions with $M_G\geq 5.5$ and XGProb\_C $\geq$0.85, we find 

627 dCs in our full sample, which is the primary, pure sample we use to derive the dC space density.  The removal of the most luminous $M_G$ bin empties the three highest $z$ bins (from 1575 to 2205\,pc), so we omit consideration of those bins below.

\begin{deluxetable}{crrrrrrrr}
\tablecaption{Visual Inspection Scores for FAST-Observed C Star Candidates \label{tab:VIscoresAll}}
\tablecolumns{9}
\tablewidth{0pt}
\tablehead{
  \colhead{VI score} &
  \colhead{$|b|>10$} &
  \colhead{a} &
  \colhead{b} &
  \colhead{c} &
  \colhead{d} &
  \colhead{e} &
  \colhead{f} &
  \colhead{g} }
\startdata 
0  &    4     &    2    &  2   &   3	& 3    &  3	    & 3	  & 3  \\
1  &    8     &    4    &  4   &   4	& 4    &  4	    & 2	  & 4  \\
2  &    81    &    14   &  15  &   15	& 21   &  15    & 9	  & 18 \\
3  &    47    &    26   &  26  &   26	& 27   &  26    & 24  & 34 \\
4  &    94    &    80   &  80  &   80	& 80   &  81    & 68  & 88 \\
5  &    331   &    318  &  325 &  327	& 326  &  328   & 313  & 324 \\
\hline
Tot &   565   &    444  &  452 & 455	& 461  &  457   & 419  & 471\\
\%$\geq$3 & 83.5 & 95.5 & 95.4 & 71.9   & 70.7 &  71.8  &  74.7 & 68.8 \\  
\%Lost & 0    &    10.2 &  8.7 &  8.3  &  8.3 &   7.8  &  14.2 &  5.5 \\    
\enddata
\tablecomments{Object counts by visual inspection score from FAST
  spectra for C star candidates in our sample, excluding those in crowded regions
  ($|b|<10\deg$ or LMC/SMC)  using 7 possible purity filters (a) through (g), which are described in \S\,\ref{sec:vi}. The row marked $\geq3$ gives the fractional purity of the candidate sample after filtering, where bona fide C stars are judged to have VI score $\geq 3$.  The final row marked \%Lost
  shows the fraction of the original uncrowded (472) candidates with VI score $\geq3$ that are lost after applying each filter.  Scoring is described in Table\,\ref{tab:viscoring}.  }
\end{deluxetable}

\begin{deluxetable}{crrrrrrrr}
\tablecaption{Visual Inspection Scores for FAST-Observed dC Candidates \label{tab:dCVIscores}}
\tablecolumns{9}
\tablewidth{0pt}
\tablehead{
  \colhead{VI score} &
  \colhead{$|b|>10$} &
  \colhead{a} &
  \colhead{b} &
  \colhead{c} &
  \colhead{d} &
  \colhead{e} &
  \colhead{f} &
  \colhead{g} }
\startdata
0  &  1   &   0  &  0   &    0  &  0   &   0  &    0  &    0  \\
1  &  3   &   1  &  1   &    1  &  1   &   1  &    1  &    1\\
2  &  52  &   9  &  9   &    9  &  14  &   9  &    3  &    12\\
3  &  12  &   6  &  6   &    6  &  6   &   6  &    5  &    8\\
4  &  19  &   17 &  17  &    17  &  17 &   17 &   10  &    17\\
5  &  54  &   53 &  53  &    53  &  53 &   53 &   51  &    54\\
\hline
Tot&  141 &   86 &  86  &    86  &  91 &   86 &   70  &    92\\
\%$\geq$ 3 &	60.3 & 88.4 & 88.4 & 88.4&	83.5 & 88.4 & 94.3 & 85.9\\  
\%Lost &  0 & 10.6 & 10.6  & 10.6  & 10.6  & 10.6  & 22.4  & 7.1  \\  
\enddata
\tablecomments{Object counts by visual inspection score from FAST
  spectra for dC candidates in our sample, excluding those in crowded regions
  ($|b|<10\deg$ or LMC/SMC)  using 7 possible purity filters (a)
  through (g), which are described in \S\,\ref{sec:vi}. The row marked $\geq3$ gives the fractional purity of the candidate sample after filtering, where bona fide C stars are judged to have VI score $\geq 3$.  The final row marked \%Lost shows the fraction of  the original uncrowded (85) candidates with VI score $\geq3$ that are lost after applying each filter.  Scoring is described in Table\,\ref{tab:viscoring}. }
\end{deluxetable}

\begin{deluxetable*}{cc|rrrrr|rrrrr}
\tablecaption{Purity Tests for dCs \label{tab:dCPurityTests}}
\tablecolumns{12}
\tablewidth{0pt}
\tablehead{
  \colhead{$M_G^{Brt}$} &
  \colhead{$M_G^{Fnt}$} &
  \multicolumn{5}{c}{No Purity Filter } &
  \multicolumn{5}{c}{Purity Filter (a)} \\
  \colhead{} & 
  \colhead{} & 
  \colhead{All} & 
  \colhead{VI $\geq$ 4} & 
  \colhead{P} & 
  \colhead{VI $\geq$ 3} & 
  \colhead{P} & 
  \colhead{All} & 
  \colhead{VI $\geq$ 4} & 
  \colhead{P} & 
  \colhead{VI $\geq$ 3} & 
  \colhead{P}
}
\startdata
4.5 & 5.5 & 35 & 3 & 0.085 & 4 & 0.1143 & 9 & 2 & 0.222 & 3 & 0.333 \\
5.5 & 6.5 & 37 & 10 & 0.270 & 18 & 0.4865 & 12 & 9 & 0.750 & 11 & 0.917 \\
6.5 & 7.5 & 44 & 35 & 0.795 & 38 & 0.8636 & 40 & 34 & 0.850 & 37 & 0.925 \\
7.5 & 8.5 & 15 & 15 & 1.000 & 15 & 1.000 & 15 & 15 & 1.000 & 15 & 1.000 \\
8.5 & 9.5 & 10 & 10 & 1.000 & 10 & 1.000 & 10 & 10 & 1.000 & 10 & 1.000 \\
\tableline
$\geq$4.5  & . & 141 & 73 & 0.518 & 85 & 0.603 & 86 & 70 & 0.814 & 76 & 0.884 \\
$\geq$5.5  & . & 106 & 70 & 0.660 & 81 & 0.764 & 77 & 68 & 0.883 & 73 & 0.948 \\
\enddata
\tablecomments{Sample purity based on visual inspection scores for dCs ($M_G\geq 4.5$) 
with FAST spectra in our candidate sample, with or without pre-filter.
Object counts in 5 bins of absolute magnitude for FAST-observed candidates, 
counting All, VI score$\geq 4$ or $\geq 3$, without any purity filter on the left, or purity filter (a) on the right. P is purity, the ratio of those that pass the VI threshold to the total. VI scoring is described in Table\,\ref{tab:viscoring}.}
\end{deluxetable*}

\section{Luminosity Function and Space Densities  \label{sec:lfsd}}

 With this new all-sky sample of C stars, we can, for the first time, measure an unbiased luminosity function and calculate the space density of dC stars. These measurements will significantly enhance our understanding of the population of dC stars and potentially provide insights into their formation.

We corrected for extinction using the \texttt{MWDUST} code \citep{Bovy2016}, using the `\texttt{combined19}' option which combines the maps of \citet{Drimmel2003}, \citet{Marshall2006}, and \citet{Green2019}. For each of our objects, we input the value of Galactic latitude ($b$) and longitude ($l$) from the Gaia DR3 data, as well as the distance estimates from \citet{Bailer-Jones2021} to the \texttt{MWDUST} code, which returned the total color excess $E(B - V)$ for each object in the sample. We calculated the extinction in the Gaia filters using Equation 2 of \citet{Canbay2023}, who calculated the extinction coefficients using Equations 1, 3a, and 3b of \citet{Cardelli1989} and assumed a standard $R_V = 3.1$. We relist the equations found in \citet{Canbay2023} below for clarity

\begin{align}
    A_G &= 0.83627 \times 3.1 E(B - V ), \\
    A_{BP} &= 1.08337 \times 3.1 E(B - V ), \\
    A_{RP} &= 0.63439 \times 3.1 E(B - V ).
\end{align}

Before we calculated and fit different models to the space density of dC stars, we applied a set of filters to ensure that our sample is a balance of pure and complete. The first filter is to ensure we only have truly \textit{dwarf} carbon stars, i.e. main-sequence stars. We do this by requiring $4.5 \leq $ M$_G \leq 9.5$ after extinction correction. The bright absolute magnitude cut is to ensure we select only main-sequence stars. The faint cut is the absolute magnitude beyond which we detect only a handful of dCs, given our apparent magnitude limit of $G=16.5$.  For highest purity and completeness, we use purity filter (a) and restrict to $M_G>5.5$, yielding a sample of 627 dCs for this analysis.

We then placed three spatial cuts: (1) removed objects in the Galactic plane with $|b| < 10^\circ$, (2) removed objects within $11^\circ$ of the LMC center, and (3) removed objects within $9^\circ$ of the SMC center. These cuts were used to remove objects likely to be affected by crowding or severe reddening in these dense regions. 

We split the sample into five bins in M$_G$, from 5.5 to 9.5 in bin widths of $0.5$\,mag. We also split the sample along the disk height, $z$, in 7 bins from $0$\,pc to $2205$\,pc in $z$ bin widths of $315$\,pc. This bin width was chosen by varying the bin width from as small as $100$\,pc to as large as $750$\,pc, and looking at the resulting luminosity function. The chosen bin width of $315$\,pc gave the best combination of $z$ resolution while ensuring all bins had enough objects to allow a statistically significant measurement of the space density. 

We then constructed the luminosity function by measuring the density of dCs in each of the bins mentioned above. For each bin in the 2D M$_G$-$z$ space, we looped through all stars in that bin. For each star in the bin we calculated the maximum distance we could see the star given the Gaia limit of 16.5 mag using 
\begin{equation}\label{eq:max_dist}
    d_{max} = 10^{\frac{16.5 - m_g + 5}{5}}.
\end{equation}
This distance is used to calculate the volume in this $z$ bin within which this star is counted. Geometrically, this volume is a slab of a sphere of radius $d_{max}$, parallel to the galactic plane, with varying portions removed based on our previous cuts. First, the volume of a slab of a sphere is given by Equation \ref{eq:slab_v}, where $h$ is the height of the slab, $a$ is the radius of the lower circular boundary of the slab, and $b$ is the radius of the upper circular boundary of the slab.

\begin{equation}\label{eq:slab_v}
    V_{slab} = \frac{1}{6} \pi h \left( 3a^2 + 3b^2 + h^2 \right)
\end{equation}

The height $h$ of the slab can be found from $z_1$ and $z_2$, which are the Galactic disk heights of our $z$ bins ($z_1$ is the lower bin edge, and $z_2$ is the upper bin edge). However, for some stars their max distance $d_{max}$ is closer than the upper bin edge, and therefore the volume is limited by Gaia and not the bin edge. In those cases ($z_2 > d_{max}$) we then set $z_2 = d_{max}$. With that adjustment, the slab height is just then $h = z_2 - z_1$.

The radii of the lower and upper circular boundaries can then be found geometrically as follows for $a$ in Equation \ref{eq:slab_a} and for $b$ in Equation \ref{eq:slab_b}. Note that for $b$, we must test if the slab height $h$ is 0, i.e. when we reach the top of the sphere and the slab has only a bottom edge.

\begin{equation}\label{eq:slab_a}
     a = \sqrt{d_{max}^2 - z_1^2}
\end{equation}

\begin{equation}\label{eq:slab_b}
    b = 
\begin{cases}
    0, & \text{if } z_1 + h = d_{max}\\
    \sqrt{d_{max}^2 - (z_1 + h)^2},              & \text{otherwise}
\end{cases}
\end{equation}

With the parameters of the slab determined, we then can calculate the slab volume as shown in Equation \ref{eq:slab_v}. However, the volume for each of our stars is not a simple slab of this sphere. We must remove the volumes of each of the spatial cuts that we placed on our sample. These are the Galactic latitude cut and the LMC and SMC cuts. Each of these cuts requires a geometric correction  to remove the volume corresponding to the cut, which is different for each slab. 

With these corrections taken into account, for each star in the M$_G$-$z$ bin, we have the calculated volume, $V_*$, that the star samples at the measured apparent magnitude, within the slab in which it is found. We then simply account for this star in the space density for each bin by adding $1 / V_*$. As we loop over all M$_G$ and $z$ bins, we account for the density of all dCs in our sample. 

Not all of our M$_G$-$z$ bins contained a dC, and therefore some of our bins do not have space density estimates. However, the bin in $z$ closest to the Galactic plane does contain objects in all the M$_G$ bins. We used the luminosity function of this $z$ bin to calculate the fraction that each M$_G$ bin contributes to the total density of each $z$ bin. This allowed us effectively to use the shape of the luminosity function to account for undetected stars in the farther $z$ bins, i.e. account for the fainter objects that are assumed to be in those more distant $z$ bins but cannot be seen due to the Gaia limit of 16.5 for XP spectra. 

We performed this correction by adding up the total space density in the closest $z$ bin, and then dividing the space density in each $M_G$ bin by the total density, giving the fraction of the total density in each M$_G$ bin. Then, for the next more distant $z$ bin, we calculated the total space density of the populated bins (note that not all bins are filled). We estimate the space density of unpopulated M$_G$ bins using the corresponding fraction from the closer $z$ bin, but normalized using the populated bins. Knowing which $M_G$ bins were populated, and the fraction of each $M_G$ bin toward the total, we calculated the corrected total density for that $z$ bin. We repeat this for each of the $z$ bins moving away from the Galactic plane, until we have filled in the density for all missing bins, resulting in the complete luminosity function for our dC sample.  While this procedure assumes that the shape of the dC luminosity function does not change significantly with $z$, it allows us to derive a more reliable space density using all the dCs available to our sample magnitude limit of $G=16.5$.  Naturally, the normalization of the luminosity function decreases with $z$, but we present evidence below (e.g., Figure\,\ref{fig:lum_func}) that its shape indeed remains substantially constant.

\begin{figure}
\epsscale{1.2}
\plotone{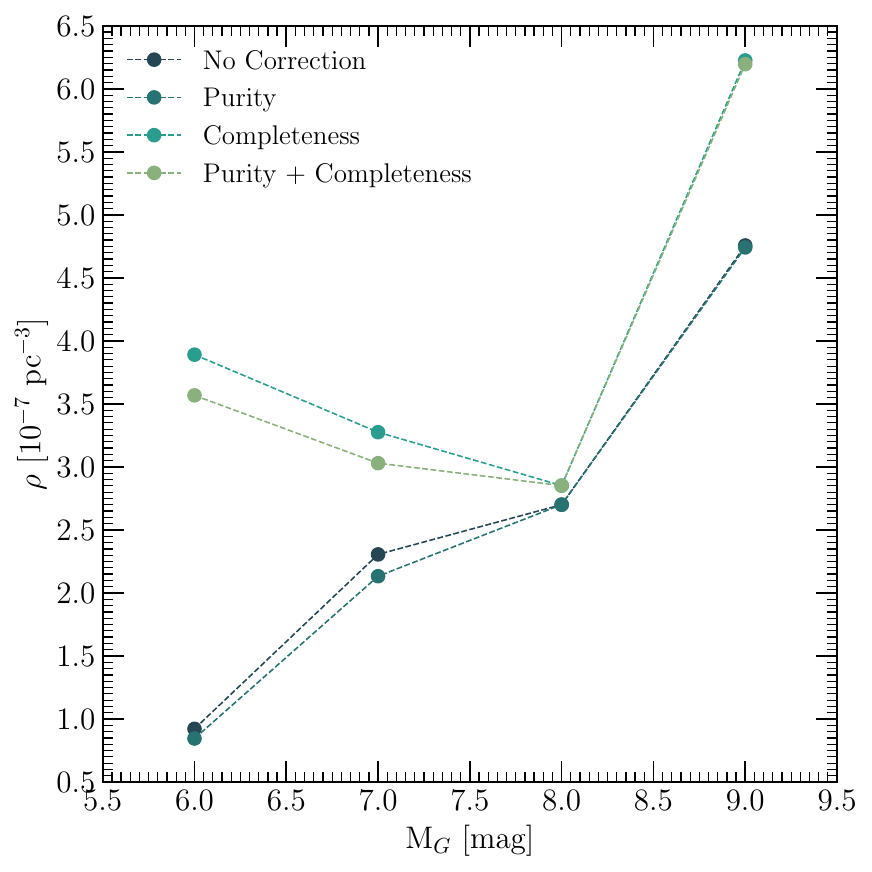}
\caption{The luminosity function for the closest $z$ bin, shown also with corrections for purity, for completeness, or both. We use both for further calculations.}
\label{fig:corr_compare}
\end{figure}

\begin{figure}
\epsscale{1.2}
\plotone{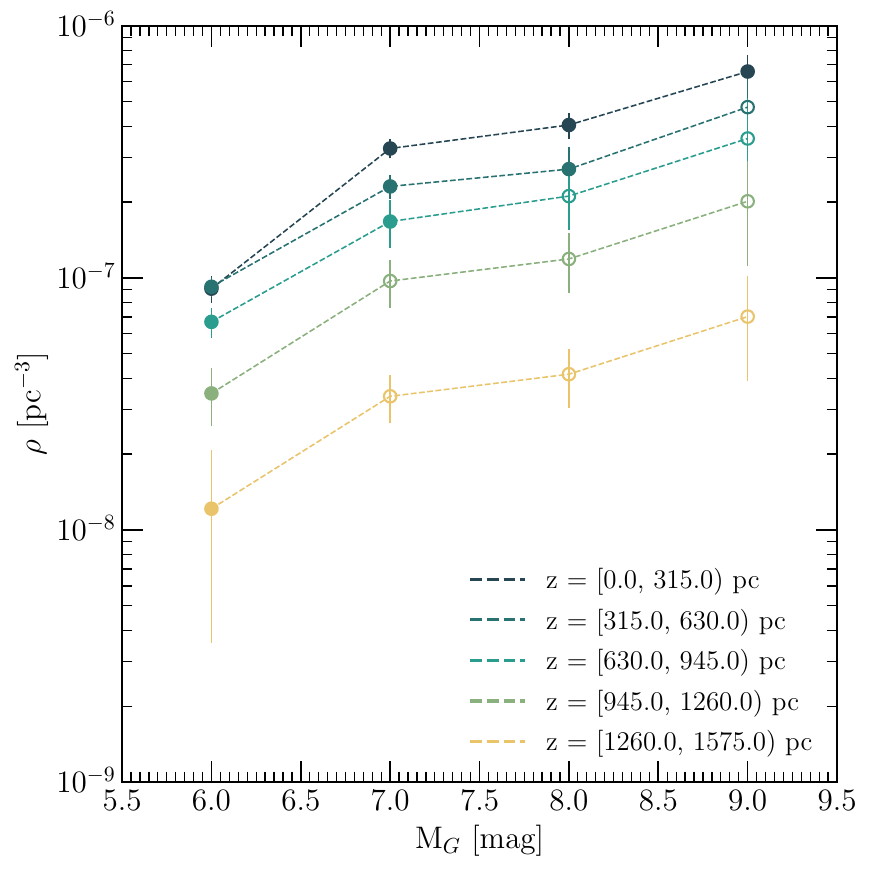}
\caption{Luminosity function, as a function of $z$ bins, of our Gaia DR3 dC candidate sample. For each bin in the disk height $z$, we calculate the space density for each bin in absolute magnitude, down to the survey limit of M$_G =$ 9.5 mag. Solid filled circles are calculated directly from the sample, with their associated error bars. Open circles are data points which have been estimated based on the measured data, as discussed in the text. The data in this figure show the space densities corrected for purity and completeness, which we use in the rest of our analysis.}
\label{fig:lum_func}
\end{figure}

To estimate the effects of the purity and completeness of our candidate sample, we provide four estimates of the dC space density. The first estimate includes all C star candidates found by our algorithm. We also reduce the star counts using an example purity filter alone, providing what is likely a lower limit to the dC space density.  Third, we apply the completeness corrections alone, which provides an indication of the maximum space density. Finally, we apply both our purity and completeness corrections to calculate the best estimate of the true space density of dCs.  Since our selection method retrieves few dCs in the range $4.5<M_G\leq 5.5$, the nominal completeness correction is unbounded, so we only estimate the dC space density for $M_G>5.5$. Figure \ref{fig:corr_compare} shows an example of the purity and completeness corrections that we applied for the closest $z$ bin.

The luminosity function is shown in Figure \ref{fig:lum_func}. Each $z$ bin is plotted with a different color. The space density in each M$_G$ bin is then plotted, with filled circles being bins that have dCs in them and open circles being bins who have had the space density interpolated from the other M$_G$ bins. In general, the shape of the luminosity function as measured by the bins with objects (filled circles) is the same across $z$ bins, although with differing normalization. For this reason, the interpolation from these filled bins into the non-filled bins from the closest $z$ bin should be a reasonable approximation to capture those objects too faint for Gaia to be accounted for in these bins.

From this luminosity function we calculated the space density as a function of $z$ height above the disk by summing over all M$_G$ bins in each $z$ bin, while calculating the associated errors on each. From this we fit the two most common geometric models for the density $z$-profile of the disk. The first was a simple exponential model in $z$ \citep{Pala2020, Dawson2024}, of the form 

\begin{equation}\label{eq:exp_model}
    \rho(z) = \rho_0  e^{-|z| / H_z}
\end{equation}

\noindent where $\rho(z)$ is the density at a given height $z$ (measured in parsecs), $\rho_0$ is the density at a height of $z=0$\,pc (i.e. at the mid-plane of the disk), and $H_z$ is the scale height.

The second model we used was a hyperbolic secant squared model \citep{Canbay2023, Dawson2024}, of the form

\begin{equation}\label{eq:sech2_model}
    \rho(z) = \rho_0  \text{sech}^2\left (\frac{|z|}{H_z} \right)
\end{equation}

\noindent where again $\rho(z)$ is the density at a given height $z$ (measured in parsecs), $\rho_0$ is the density at a height of $z=0$\,pc (i.e. at the mid-plane of the disk), and $H_z$ is a characteristic height of the disk.

We used the same method for both models to find the best set of parameters that fit our data. In each case we used the \citet{Goodman2010} Markov chain Monte Carlo (MCMC) ensemble sampler implemented in \texttt{emcee} \citep{emcee}. For our sampler we used 100 walkers and 110,000 iterations, removing the first 10,000 as a burn-in period. To ensure the samplers converged, we used $\ln{\rho_0}$ as the parameter sampled to avoid numerical convergence issues. For both models we used a uniform prior over both $\ln{\rho_0}$ and $H_z$. The allowed range for $\ln{\rho_0}$ was between $-20.7$ and $-7$, while the allowed range for $H_z$ was between $0$\,pc and $1500$\,pc.

\begin{figure*}
\gridline{\fig{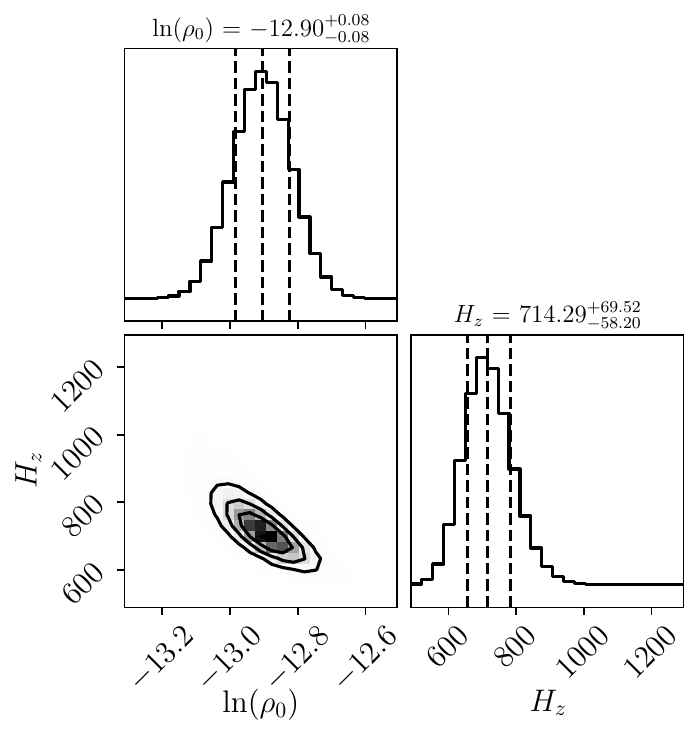}{0.5\textwidth}{(a)}
          \fig{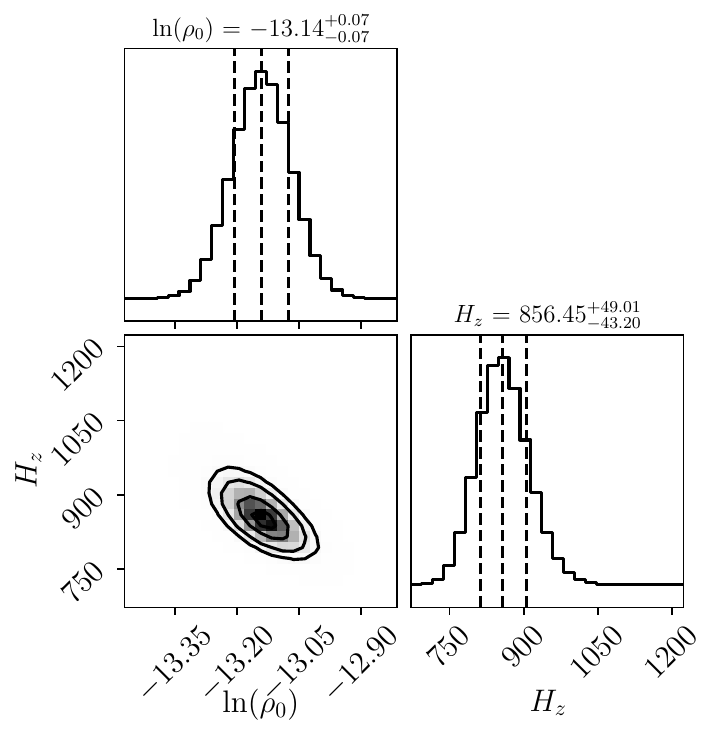}{0.5\textwidth}{(b)}}
\caption{Posterior distributions for the model parameters $\rho_0$, H$_z$, the central disk density and scale height respectively for star counts incorporating both purity and completeness corrections. Shown are the model fit results for an exponential model (a) and a hyperbolic secant model (b). Vertical dashed lines represent the 1$\sigma$ range and the median (50th percentile). Values are the natural logarithm (ln) of the central disk density in units of pc$^{-3}$.}
\label{fig:mcmc_post}
\end{figure*}

The resulting sampled posterior distributions for both models are shown in Figure \ref{fig:mcmc_post}. The left panel shows the posterior for the exponential model, while the right panel shows the hyperbolic secant squared model. In both cases the posterior distribution is shown as a 2-D density histogram with 1, 2, and 3$\sigma$ level contours, as well as marginalized posterior distributions for both of the model parameters. From these marginalized histograms we take the median as the best fit value, and report our errors as the 1$\sigma$ levels. We performed this model fit with our four different corrected space densities: no correction, purity, completeness, and purity and completeness. The results of all these fits are found in Table \ref{tab:rhoHfits}.
There seem to be only minor variations in the fit values among the various datasets and models. We present all the values in Table \ref{tab:rhoHfits}, but for our subsequent analysis, we utilize the fit including corrections for both purity and completeness.  
\begin{deluxetable}{ccccc}
\tablecaption{Space Density and Scale Height Fit Results \label{tab:rhoHfits}}
\tablehead{
\colhead{Correction} & \colhead{Model} & \colhead{$\rho_0$}  & \colhead{$H_Z$} & \colhead{BIC} \\
\colhead{} & \colhead{} & \colhead{$10^{-6}$\,pc$^{-3}$}  & \colhead{pc} &  
}
\startdata
None    & Exp        & $1.83^{+0.16}_{-0.16}$ & $642^{+60}_{-51}$ & $2.04$  \\
\ldots  & sech($Z$)  & $1.41^{+0.11}_{-0.10}$ & $819^{+47}_{-41}$  & $-3.86$  \\
Purity  & Exp        & $1.78^{+0.17}_{-0.16}$ & $641^{+62}_{-52}$ & $1.77$  \\
\ldots  & sech($Z$)  & $1.38^{+0.11}_{-0.10}$ & $818^{+48}_{-42}$  & $-3.71$  \\
Completeness   & Exp & $2.57^{+0.21}_{-0.19}$ & $724^{+71}_{-59}$ & $3.61$  \\
\ldots  & sech($Z$)  & $2.04^{+0.13}_{-0.13}$ & $861^{+49}_{-43}$  & $-4.75$  \\
P$+$C   & Exp        & $2.49^{+0.20}_{-0.19}$ & $714^{+70}_{-58}$ & $3.50$  \\
\ldots  & sech($Z$)  & $1.96^{+0.14}_{-0.12}$ & $856^{+49}_{-43}$  & $-4.57$   \\
\enddata
\tablecomments{Best-fit model parameters for single exponential or hyperbolic secant density distribution as a function of height $z$ from the mid-plane.  The sample modeled here includes only dCs in the range of dereddened absolute magnitudes $5.5<M_{G,0}<9.5$. The errors reported are standard asymmetric $1\sigma$ confidence interval from the MCMC samples. }
\end{deluxetable}

\begin{figure}
\epsscale{1.2}
\plotone{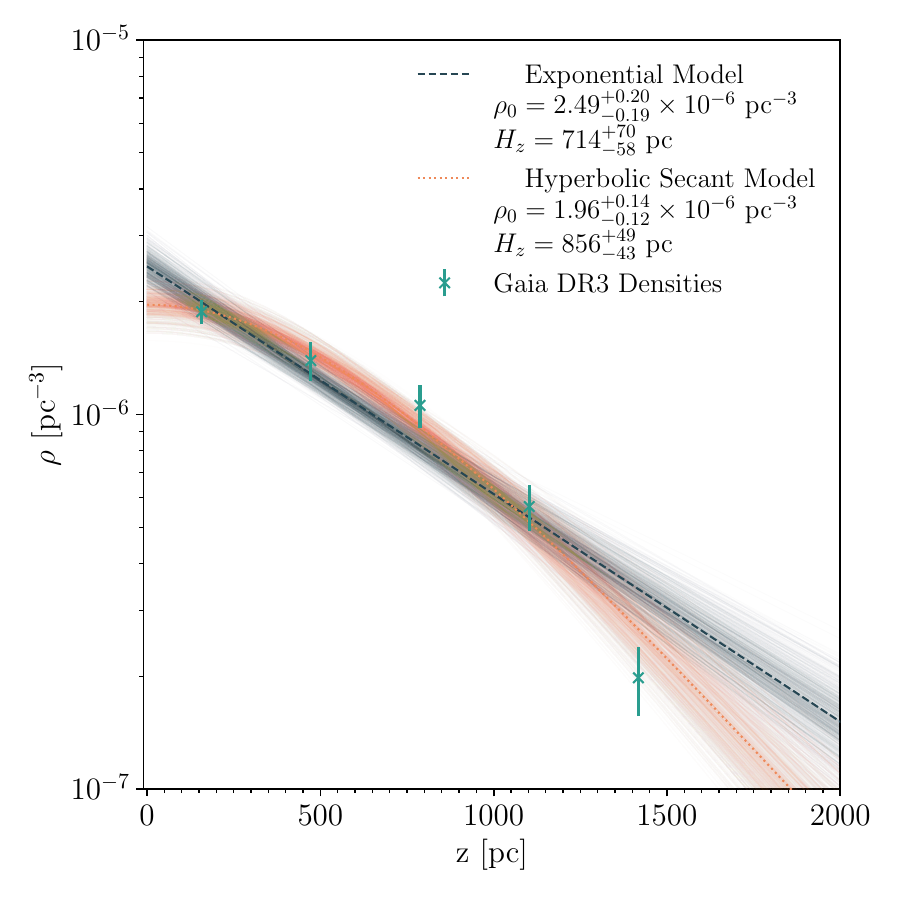}
\caption{Disk model fits to the density profiles of the dC candidate sample. Each data point in a $z$ bin represents the total space density across all absolute magnitudes. We fit both an exponential model (in dark blue) and a hyperbolic secant model (in orange). For both models, we show the median of each distribution (dashed for exponential, dotted for hyperbolic secant) as the model in the darker line. Additionally, we plot 1000 random draws from each posterior distribution to illustrate the range of possible models. Based on the data, the hyperbolic secant model appears to be a better fit to the observed density profiles.}
\label{fig:density_fit_plot}
\end{figure}

In Figure \ref{fig:density_fit_plot}, we present the total measured space density as a function of our $z$ bins, along with their associated errors. Both models are depicted with a line (dashed or dotted), indicating the model derived from the optimized parameters. Moreover, we also plot, in transparent lines of the same color, 1000 randomly generated models drawn from the posterior distribution of the model fits. This visualization illustrates the range of models that are probable around the best-fit median model.

To determine which model better fits the observed space densities from Gaia, we calculated the Bayesian information criterion (BIC) for each model and reported the results in Table \ref{tab:rhoHfits}. For all four datasets, we found that the hyperbolic secant squared model has a lower BIC. Moreover, for the purity and completeness-corrected data, this change in BIC is $\Delta \text{BIC} = 8.07$, indicating very strong evidence that the hyperbolic secant squared model more accurately represents the data. 

Consequently, we report the mid-plane space density and scale height of dC stars in the galaxy as the fit parameters derived from this hyperbolic secant squared model for the purity and completeness-corrected data. This gives us the mid-plane space density of dC stars as $\rho_0\,=\,1.96^{+0.14}_{-0.12}\times10^{-6}\,\text{pc}^{-3}$ and a scale height of $H_z\,=\,856^{+49}_{-43}\,$pc.

As regards our measured scaleheight, the value 
that we find for dCs of $H_z\sim\,800\,$\,pc is surprisingly large compared to G and K dwarfs ($\sim 150$pc; \citealt{Bovy2017}).  Velocity dispersion increases with the age of stars due to dynamical heating of the disk (e.g, \citealt{Nordstrom2004, Aumer2009} and references therein).  As a result, higher mass stars, which have shorter lifetimes tend to have lower velocities and remain closer to the plane of the galaxy.  Older stars with lower masses have larger scaleheights.  It is possible that dCs are principally from an older population, given that each has a companion that already went through an AGB phase.  Furthermore, it takes less C-enriched mass transfer to create a (C$>$O) dC if the system is low metallicity.  Low metallicity corresponds to greater age, which in turn corresponds to larger scaleheights (e.g., \citealt{Gilmore1983, Bovy2012}).

In the following section, we compare the space density of dCs to other relevant stellar populations.

\section{Space Density Comparisons  \label{sec:compareSD}}

We present in this article the first uniformly-selected all-sky catalog of carbon star candidates.  We measure the completeness and purity of the candidate sample, and focus our attention on the dwarf carbon stars.  The subsample with the highest purity and completeness are the cooler dCs, with $M_G>5.5$.  We can contrast the dC space density $\rhodC \sim 2\times 10^{-6}$\,pc$^{-3}$ at the mid-plane to that of other stellar types in the literature.  For perspective, remember that the preponderance of dCs in our sample are in binary star systems where (1) the initial primary star has been through the TP-AGB phase to become an AGB C star, (2) significant mass transfer has occurred to tip C$>$O in the atmosphere of the remaining main sequence star, (3) the final dC mass corresponds to an effective temperature cool enough to allow the formation of detectable C$_2$ and/or CN molecular bands and (4) the WD remnant has cooled beyond obvious detectability ($T_{WD}\leq$ 10,000 deg).

\subsection{Main Sequence Stars \label{sec:compareMS}}

\citet{Bovy2017} measured the space density of A0V through K4V main sequence stars using about two million stars in the Gaia DR1 {\em TGAS} catalog.  We can compare our space density \rhodC\, to theirs in a similar absolute magnitude range.  We calculate space density for $5.5<M_G<9.5$ which corresponds (for dCs) to $4.0<M_J<7.6$.  Using the mean infrared dwarf stellar locus from \citet{Pecaut2013}, a similar range in $M_J$ corresponds to $\geq$G8V for normal main sequence stars.  Bovy only tabulates up through K4V ($M_J=7.25$) because of incompleteness, which corresponds to $M_G=9.3$ or so for dCs.  Their cumulative space density fom G8V to K4V, $5.20\times 10^{-3}$ pc-3,  is some 2600 times larger than we find for \rhodC\, in a similar $M_J$ range.  Therefore, local dCs represent only about 0.03\% of main sequence stars, as might be expected because of dCs' very particular evolutionary history.  

We note that dC stars have more similarity in the infrared to normal (C$<O$) main sequence stars than in the optical.  The trend of $M_K$ with stellar bolometric luminosity is quite similar.  If $M_K$ is a good proxy for stellar mass, then our dCs range in mass from about 0.1 to 0.55\,\Msun.  

\subsection{AGB Carbon Stars \label{sec:compareAGBC}}

By comparing the space density of dCs to that of C-AGB stars, we can crudely estimate the fraction of C-AGB stars in systems that produce dCs.  Since dCs may remain on the main sequence for billions of years after the mass transfer event from their former AGB companion, the dC population effectively registers many generations of past C-AGB stars.  The scale height of dCs that we measure indicates thick disk kinematics.  Thick disk populations have a typical median age of about 10Gyr (e.g., \citealt{Haywood2013}). The duration of the TP-AGB phase generally decreases with increasing mass or metallicity \citep{Girardi2007, Kalirai2014}, but we can assume for simplicity that a typical AGB C star lifetime is about 1Myr.  If all AGB C stars birthed a dC, dCs should thus have a space density some 10$^4$ times greater than C-AGB stars. In fact the dC space density is only about 200 times higher.  This would imply that only about $\sim$2\% of C-AGB stars are in mass transfer binary systems that result in a detectable dC star.  Why only 2\%?  The C-AGB star had to be in a binary system.  The system had to be close enough for mass transfer, but not so close as to quickly disrupt the TP-AGB phase or to merge with the AGB primary.  The initial mass ratio had to be large enough that the dC star remains on the main sequence, with a mass low enough that we can detect C$_2$ and CN features.  In upcoming work, we will refine the space density measurement for C-AGB stars using our sample, and further investigate these arguments.
\\

\subsection{White Dwarfs \label{sec:compareWD}}

WDs are created after either a red giant or an AGB phase. The space density of single WDs should be considerably larger than dCs because they need not have been C-AGB stars, nor have main sequence companions.    \citet{GentileFusillo2021} inferred a space density of white dwarfs within 20\,pc of the Sun of 4.47$\pm 10^{-3}$\,pc$^{-3}$, using the Gaia EDR3 catalogue. A more recent estimate using 1076 spectroscopically confirmed white dwarfs within 40pc from Gaia DR3 yields a consistent space density \citep{OBrien2024}.

Therefore, detectable WDs are at least 2200 times as common as dCs.   Detectability eliminates the coolest, oldest WDs, so this measured $\rho_{\rm WD}$ value is certainly a lower limit on the total intrinsic space density of WDs.

\subsection{Unresolved White Dwarf/Main Sequence Binary Stars  \label{sec:compareWDMS}}

In terms of stellar evolution, the most closely analogous systems to dCs are likely the WDMS binaries.   Using Gaia EDR3, \citet{Rebassa-Mansergas2021} select 112 WDMS binaries within 100\,pc from a region of the color-magnitude diagram between the WD and main sequence.  They derive  space density $\rho\sim 3\times 10^{-5}$\,pc$^{-3}$.  However, detection either via color or spectroscopy requires that the WD and its dwarf companion have similar absolute magnitudes.  Thus, the WD must still be hot, especially for those systems with more luminous (more massive) main sequence companions. Based on binary population simulations, they estimate that their incompleteness is about 90\%, because in the intrinsic population, the WD is usually much less luminous than its MS companion and therefore undetectable. Correction for this severe incompleteness yields an estimated upper limit to the true WDMS space density of $\sim 4\times 10^{-4}$\,pc$^{-3}$, or about 200 times the dC space density.  This much larger value is due in large part to the narrower evolutionary path of dCs; while the WD in WDMS systems was surely a giant, the initial primary in a dC system must have undergone the TP-AGB to become an AGB C star and also transferred substantial mass to the current dC, without having subsumed the dC in the process.

Interestingly, we may be able to improve the comparison of WDMS and dC space densities using the relative frequency of C-AGB to M-AGB stars, albeit with some unjustified assumptions. 

In M31, for solar metallicity (or 0.1 solar), the C/M ratio is about 0.01 (0.05) \citep{Boyer2019}.  

If the WDs in WDMS systems had AGB progenitors with similar C/M ratios, we might expect the intrinsic space density of WDMS systems to thus be 20 - 100 times that of dCs.
However, the hot WD in a typical WDMS system could have come from either a red giant or an AGB star, and so the WD progenitor mass range is quite large, from about 1 -- 8\Msun\  \citep{Cummings2018}.  In comparison, the progenitor mass range of C-AGB stars is likely just $\sim 1.5 - 2.5$\Msun, with some dependence on metallicity \citep{Straniero2023}. This difference in mass range, integrated over a basic initial mass function (e.g., $m^{-2.3}$; \citealt{Kroupa2001}) boosts the expected ratio of WDMS vs. dC space densities by about another factor of 3.2, which may account for the observed ratio of $\sim$200, within the still significant uncertainties.

\subsection{Binary Population Synthesis  \label{sec:BPS}}

Unlike WDMS systems, WD or main sequence samples, we know that every dC has a WD companion, whether or not we  detect it. The comparisons above merely highlight that the space density of dCs is best compared to results from binary population synthesis models, because dCs provide a unique set of constraints.  The necessary components of applicable binary evolution models include assumptions about the binary fraction, the mass ratio distribution, mass transfer, wind loss, angular-momentum transfer, and common-envelope evolution. Many details of these calculations are under debate including e.g., the duration of the CE phase, which is especially important for dCs with short period orbits \citet{Roulston2021}, but could range from just 100 days to 10$^4$ years (e.g., \citealt{Igoshev2020}). The time-integrated space density of C-AGB stars also comes into play, which depends on detailed stellar evolution models including such parameters as the initial mass-range, metallicity effects, the amount of convective overshoot, etc. (e.g., \citealt{Stancliffe2005}). 

Comparison to measured space densities also requires realistic application of selection criteria to the simulated population.  
\citet{Kool1995} made the earliest theoretical estimate of the space density of dCs, predicting
a local space density within the disk of $1 \times 10^{-6} \,\textrm{pc}^{-3}$, which is just a factor of two lower than our measurement.  

The comparison sample at the time was based on just a few high proper motion objects detected among faint high latitude halo C stars in objective prism surveys \citep{Green1992} and a handful of DA/dCs with disk-like kinematics \citep{Heber1993,Liebert1994}.   The current work provides a much more reliable estimate of the local space density of dCs.  
Significant advances in binary population synthesis and in our understanding and modeling of AGB evolution and mass transfer suggest that new comparisons between predictions and observations would be productive.

\section{Summary and Prospects  \label{sec:summary}}

By using XP spectra from {\em Gaia} DR3, and training machine learning algorithms on a sample of carbon stars verified with higher resolution spectroscopy, we have created by far the largest catalog of carbon stars ever made, across the entire sky.  We measure the purity and completeness of the sample, and derive the local mid-plane space density from 626 dCs with $5.5<M_G<9.5$ having XGProb\_C value above 0.85, a sample with 94.8\% purity.  The dCs have a space density of $\rho_0\,\sim 2\times10^{-6}\,\text{pc}^{-3}$ and a scale height of $H_z\,\sim\,850\,$pc.

High resolution spectroscopy of dCs is rare to date, and so verified model atmospheres for dCs do not exist. Both are important, because dCs \citep{Plez2005} and their evolved descendants such as CEMP-$s$ stars (e.g.,, \citealt{Beers2005}), have been posited as some of the most metal-poor stars in the galaxy.  Restricting to uncrowded regions, XGProb\_C$>0.85$, and $M_G>5.5$, there are now 24 dCs in our catalog with $G<14$.  Only two are previously known as dCs - the prototype dC G77-61 \citep{Dahn1977} at $G=13.32$ and LP 318-342 \citep{Si2015} at $G=13.43$.  We have discovered four dCs even brighter than those, with two having $G=12.5$.  If we don’t restrict to $|b|>10^{\circ}$, there are 31 dCs with $G<14$.  These bright dCs are well-suited to high resolution spectroscopy and atmospheric modeling.  

We are still obtaining spectroscopy of our dC candidates, both to confirm that they are true dC stars, and to obtain radial velocities (RVs).  For true dCs with periodic variability detected from photometric lightcurves, we obtain multi-epoch spectroscopy to measure orbital properties. For every dC we observe, we can derive space velocities from the RVs along with {\em Gaia} proper motions and parallaxes.  We will thus study the kinematics of our dC sample. Disk stars tend to concentrate near the Galactic plane ($|z| < 1-2$\,kpc) and display relatively low proper motions due to rotational alignment. Halo stars are more widely dispersed along the $z$-axis, extending farther from the plane with generally higher proper motion magnitudes.A kinematic study of dCs will help us understand their age, and relative frequency in disk and halo.

To better constrain the evolutionary path leading to dCs, it is important to know the space density of C-AGB stars. We are extending our analysis in an upcoming paper to model the distribution of C-AGB stars, which can be reliably detected out to about 50\,kpc for $G<16.5$.  

It is also key  to find the most recently-minted dCs - those where the DA WD is still hot.  Too few DAdC systems are known at bright magnitudes to form an adequate training sample, but we are synthesizing and combining spectra for dC and DA WD templates for a range of temperatures, luminosities and distances.

Large multifiber spectroscopic surveys are underway now across the full sky (e.g. WEAVE, 4MOST, DESI, SDSS-V; \citealt{4MOST, WEAVE,  DESI2016, SDSS_5}), which will provide significantly enhanced samples of carbon stars at fainter magnitudes.  As the space density of dCs is considerably higher than that of C giants, the majority of new faint C star spectra will be dCs.  At magnitudes of $G\sim$19 or fainter, AGB giants ($M_G<-2$, but TP-AGB stars have $M_G\leq -4$) must be at distances over 150\,kpc, so well outside the galactic halo.  By contrast for dCs, such faint surveys are only probing distances of about 5\,kpc.

\facility{Gaia, LAMOST}

\software{Astropy \citep{astropy1, astropy2}, Matplotlib \citep{matplotlib}, Numpy \citep{numpy}, Scipy \citep{scipy}, Scikit-Learn \citep{Scikit-learn}, TOPCAT \citep{topcat}}

\begin{acknowledgments}
B.R. was supported for this work  by the National Aeronautics and Space Administration through Chandra Award Numbers GO0-21003X and GO1-22004X issued by the Chandra X-ray Center, which is operated by the Smithsonian Astrophysical Observatory for and on behalf of the National Aeronautics Space Administration under contract NAS8-03060.  Additional support for observations made with the NASA/ESA Hubble Space Telescope was provided under program number GO-16392, provided by NASA through a grant from STScI, which is operated by AURA, Inc., under NASA contract NAS 5-26555. 

We are grateful for the Vizier service \citep{Vizier2000}, which we accessed for numerous purposes during this study.  This research made use of the SIMBAD database, operated at CDS, Strasbourg, France \citep{Simbad2000}

This study made use of data from the European Space Agency (ESA) mission {\it Gaia} (\url{https://www.cosmos.esa.int/gaia}), processed by the {\it Gaia} Data Processing and Analysis Consortium (DPAC, \url{https://www.cosmos.esa.int/web/gaia/dpac/consortium}). Funding for the DPAC
 has been provided by national institutions, in particular the institutions
 participating in the {\it Gaia} Multilateral Agreement.

This study made use of the Python package GaiaXPy, developed and maintained by members of the Gaia Data Processing and Analysis Consortium (DPAC), and in particular, Coordination Unit 5 (CU5), and the Data Processing Centre located at the Institute of Astronomy, Cambridge, UK (DPCI).

Funding for the Sloan Digital Sky Survey IV has been provided by the Alfred P. Sloan Foundation, the U.S. Department of Energy Office of Science, and the Participating Institutions. 

SDSS-IV acknowledges support and resources from the Center for High Performance Computing  at the University of Utah. The SDSS website is www.sdss.org.

SDSS-IV is managed by the Astrophysical Research Consortium for the Participating Institutions of the SDSS Collaboration including the Brazilian Participation Group, the Carnegie Institution for Science, Carnegie Mellon University, Center for Astrophysics $|$ Harvard \& Smithsonian, the Chilean Participation Group, the French Participation Group, Instituto de Astrof\'isica de Canarias, The Johns Hopkins University, Kavli Institute for the Physics and Mathematics of the Universe (IPMU) / University of Tokyo, the Korean Participation Group, Lawrence Berkeley National Laboratory, Leibniz Institut f\"ur Astrophysik Potsdam (AIP),  Max-Planck-Institut f\"ur Astronomie (MPIA Heidelberg), Max-Planck-Institut f\"ur Astrophysik (MPA Garching), Max-Planck-Institut f\"ur Extraterrestrische Physik (MPE), National Astronomical Observatories of China, New Mexico State University, New York University, University of Notre Dame, Observat\'ario Nacional / MCTI, The Ohio State University, Pennsylvania State University, Shanghai Astronomical Observatory, United Kingdom Participation Group, Universidad Nacional Aut\'onoma de M\'exico, University of Arizona, University of Colorado Boulder, University of Oxford, University of Portsmouth, University of Utah, University of Virginia, University of Washington, University of Wisconsin, Vanderbilt University, and Yale University.
\end{acknowledgments}

\appendix

\section{Initial Carbon Star Training Sample \label{sec:initCtraining}}

We compiled our first carbon star training sample from published sources where the C-star classification was based on existing optical spectroscopy.  

For our initial training sample, we selected carbon stars from SIMBAD by requiring that sptypes$=$'C*', yielding 

objects labeled as C, C-H, C-Hd, C-R, C-N, C-J, CEMP.  Many had more detailed spectral typing from the literature, but the initial sample is quite heterogeneous, based on spectral classification from low-resolution objective prism (e.g., \citealt{Christlieb2001}) to intermediate resolution SDSS amd LAMOST (e.g., \citealt{Green2013,Li2018}), to 
high resolution spectroscopy (e.g., \citealt{Cotar2019}. Overall, the initial sample (without being matched to Gaia DR3) consisted of 22312 objects. 

We eliminated C stars with Galactic longitude $|b|<10$deg, to avoid the worst effects of reddening. Then, we eliminated C stars detected in the Large and Small Magellanic Clouds, because of potential croding issues, and because luminous C giants are already well represented within the Milky Way sample. This was accomplished by using circular regions with radii of 11\,deg and 9\,deg from the centers of the LMC and SMC respectively. The galactic coordinates for these centers were found from NASA/IPAC Extragalactic Database. After this cut, our sample was reduced to 5657 stars before we cross-matched to the Gaia DR3 catalog, which yielded 

a sample of 5286 entries.
Out of those stars, only 3686 had XP spectra and G mean flux over error $>$ 5.

Both Ba and CEMP stars often lack detectable C$_2$ or CN bands, and including them in the C star training sample could result in a substantial contamination of our C star candidate sample with normal (C$<$O) stars, so we eliminated both classes (536 CEMP and 381 Ba stars) 
from our C star training sample for the sake of purity. We also identified and eliminated one known DA/dC spectroscopic binary, SDSS J081157.13+143533.0 \citep{Whitehouse2018}. We performed a visual inspection of the XP spectra of these stars, overplotting a higher resolution template C star spectrum for comparison.  This inspection revealed that many - mostly faint - stars had large, unphysical fluctuations.  We therefore chose to eliminate the remaining 233 C stars with $G>16.9$.  We further eliminated 105 stars with large discontinuities, evident TiO band absorption, or unclear C$_2$ and CN bands. Our initial C star training sample thus contained 2431 stars.

\section{Initial Control Star Training Sample  \label{sec:controlTraining}}

For each C star in our training sample, we find a random star\footnote{We used Python's np.random on the top eight million stars pulled from an ADQL query of Gaia DR3.} 

$|\Delta \overline{G}| \leq 0.1$, $|\Delta (\bprp )| \leq 0.1$

Since we do not require significant a parallax measurement either for the C stars or the control stars, this matching does not guarantee that the luminosities of a C star and its matching control star are similar.  Many of the C stars are distant giants, and have parallaxes too small to measure with Gaia's current sensitivity.
There is a fair variety in appearance of the Gaia spectra of control stars, even if they have similar colors and magnitudes.  For instance, we might expect all of the reddest control stars to have similar strength TiO bands, but this is not the case.  Possible reasons may include differences in luminosity, metal abundances, dust along the line of sight, etc.  We do not attempt to eliminate this variety among the control sample, because random field stars will have the same dispersion in properties.   A few of the initial control sample stars (31 stars) we determined through visual inspection had clear C$_2$ or CN absorption, or unphysical discontinuities. These were randomly replaced with another control star from Gaia. 

We trained XGBoost to select C star candidates (as described in \S\,\ref{sec:ML}) using this initial SIMBAD-based C star training sample, and the color-matched control sample. When we did a preliminary inspection of FAST spectra from the resulting C star candidate lists, we obtained estimates of only 60\% purity. We speculate the low purity may be due to some combination of the heterogeneity of the C star training sample and the relatively small size and breadth of the color-matched control sample. As constructed, the control sample would not effectively filter out objects like galaxies and quasars from carbon star spectra. For these reasons, we chose to use the LAMOST C star training sample and  the much larger extended control sample, as described in \S\,\ref{sec:cstarTraining}.

\bibliography{main}{}

\begin{thebibliography}{}
\expandafter\ifx\csname natexlab\endcsname\relax\def\natexlab#1{#1}\fi
\providecommand{\url}[1]{\href{#1}{#1}}
\providecommand{\dodoi}[1]{doi:~\href{http://doi.org/#1}{\nolinkurl{#1}}}
\providecommand{\doeprint}[1]{\href{http://ascl.net/#1}{\nolinkurl{http://ascl.net/#1}}}
\providecommand{\doarXiv}[1]{\href{https://arxiv.org/abs/#1}{\nolinkurl{https://arxiv.org/abs/#1}}}

\bibitem[{{Abia} {et~al.}(2022){Abia}, {de Laverny}, {Romero-G{\'o}mez}, \&
  {Figueras}}]{Abia2022}
{Abia}, C., {de Laverny}, P., {Romero-G{\'o}mez}, M., \& {Figueras}, F. 2022,
  \aap, 664, A45, \dodoi{10.1051/0004-6361/202243595}

\bibitem[{{Astropy Collaboration} {et~al.}(2013){Astropy Collaboration},
  {Robitaille}, {Tollerud}, {Greenfield}, {Droettboom}, {Bray}, {Aldcroft},
  {Davis}, {Ginsburg}, {Price-Whelan}, {Kerzendorf}, {Conley}, {Crighton},
  {Barbary}, {Muna}, {Ferguson}, {Grollier}, {Parikh}, {Nair}, {Unther},
  {Deil}, {Woillez}, {Conseil}, {Kramer}, {Turner}, {Singer}, {Fox}, {Weaver},
  {Zabalza}, {Edwards}, {Azalee Bostroem}, {Burke}, {Casey}, {Crawford},
  {Dencheva}, {Ely}, {Jenness}, {Labrie}, {Lim}, {Pierfederici}, {Pontzen},
  {Ptak}, {Refsdal}, {Servillat}, \& {Streicher}}]{astropy1}
{Astropy Collaboration}, {Robitaille}, T.~P., {Tollerud}, E.~J., {et~al.} 2013,
  \aap, 558, A33, \dodoi{10.1051/0004-6361/201322068}

\bibitem[{{Astropy Collaboration} {et~al.}(2018){Astropy Collaboration},
  {Price-Whelan}, {Sip{\H{o}}cz}, {G{\"u}nther}, {Lim}, {Crawford}, {Conseil},
  {Shupe}, {Craig}, {Dencheva}, {Ginsburg}, {VanderPlas}, {Bradley},
  {P{\'e}rez-Su{\'a}rez}, {de Val-Borro}, {Aldcroft}, {Cruz}, {Robitaille},
  {Tollerud}, {Ardelean}, {Babej}, {Bach}, {Bachetti}, {Bakanov}, {Bamford},
  {Barentsen}, {Barmby}, {Baumbach}, {Berry}, {Biscani}, {Boquien}, {Bostroem},
  {Bouma}, {Brammer}, {Bray}, {Breytenbach}, {Buddelmeijer}, {Burke},
  {Calderone}, {Cano Rodr{\'\i}guez}, {Cara}, {Cardoso}, {Cheedella}, {Copin},
  {Corrales}, {Crichton}, {D'Avella}, {Deil}, {Depagne}, {Dietrich}, {Donath},
  {Droettboom}, {Earl}, {Erben}, {Fabbro}, {Ferreira}, {Finethy}, {Fox},
  {Garrison}, {Gibbons}, {Goldstein}, {Gommers}, {Greco}, {Greenfield},
  {Groener}, {Grollier}, {Hagen}, {Hirst}, {Homeier}, {Horton}, {Hosseinzadeh},
  {Hu}, {Hunkeler}, {Ivezi{\'c}}, {Jain}, {Jenness}, {Kanarek}, {Kendrew},
  {Kern}, {Kerzendorf}, {Khvalko}, {King}, {Kirkby}, {Kulkarni}, {Kumar},
  {Lee}, {Lenz}, {Littlefair}, {Ma}, {Macleod}, {Mastropietro}, {McCully},
  {Montagnac}, {Morris}, {Mueller}, {Mumford}, {Muna}, {Murphy}, {Nelson},
  {Nguyen}, {Ninan}, {N{\"o}the}, {Ogaz}, {Oh}, {Parejko}, {Parley}, {Pascual},
  {Patil}, {Patil}, {Plunkett}, {Prochaska}, {Rastogi}, {Reddy Janga},
  {Sabater}, {Sakurikar}, {Seifert}, {Sherbert}, {Sherwood-Taylor}, {Shih},
  {Sick}, {Silbiger}, {Singanamalla}, {Singer}, {Sladen}, {Sooley},
  {Sornarajah}, {Streicher}, {Teuben}, {Thomas}, {Tremblay}, {Turner},
  {Terr{\'o}n}, {van Kerkwijk}, {de la Vega}, {Watkins}, {Weaver}, {Whitmore},
  {Woillez}, {Zabalza}, \& {Astropy Contributors}}]{astropy2}
{Astropy Collaboration}, {Price-Whelan}, A.~M., {Sip{\H{o}}cz}, B.~M., {et~al.}
  2018, \aj, 156, 123, \dodoi{10.3847/1538-3881/aabc4f}

\bibitem[{{Aumer} \& {Binney}(2009)}]{Aumer2009}
{Aumer}, M., \& {Binney}, J.~J. 2009, \mnras, 397, 1286,
  \dodoi{10.1111/j.1365-2966.2009.15053.x}

\bibitem[{{Bailer-Jones} {et~al.}(2021){Bailer-Jones}, {Rybizki}, {Fouesneau},
  {Demleitner}, \& {Andrae}}]{Bailer-Jones2021}
{Bailer-Jones}, C.~A.~L., {Rybizki}, J., {Fouesneau}, M., {Demleitner}, M., \&
  {Andrae}, R. 2021, \aj, 161, 147, \dodoi{10.3847/1538-3881/abd806}

\bibitem[{{Beers} \& {Christlieb}(2005)}]{Beers2005}
{Beers}, T.~C., \& {Christlieb}, N. 2005, \araa, 43, 531,
  \dodoi{10.1146/annurev.astro.42.053102.134057}

\bibitem[{{Bellm} {et~al.}(2019){Bellm}, {Kulkarni}, {Graham}, {Dekany},
  {Smith}, {Riddle}, {Masci}, {Helou}, {Prince}, {Adams}, {Barbarino},
  {Barlow}, {Bauer}, {Beck}, {Belicki}, {Biswas}, {Blagorodnova}, {Bodewits},
  {Bolin}, {Brinnel}, {Brooke}, {Bue}, {Bulla}, {Burruss}, {Cenko}, {Chang},
  {Connolly}, {Coughlin}, {Cromer}, {Cunningham}, {De}, {Delacroix}, {Desai},
  {Duev}, {Eadie}, {Farnham}, {Feeney}, {Feindt}, {Flynn}, {Franckowiak},
  {Frederick}, {Fremling}, {Gal-Yam}, {Gezari}, {Giomi}, {Goldstein},
  {Golkhou}, {Goobar}, {Groom}, {Hacopians}, {Hale}, {Henning}, {Ho}, {Hover},
  {Howell}, {Hung}, {Huppenkothen}, {Imel}, {Ip}, {Ivezi{\'c}}, {Jackson},
  {Jones}, {Juric}, {Kasliwal}, {Kaspi}, {Kaye}, {Kelley}, {Kowalski},
  {Kramer}, {Kupfer}, {Landry}, {Laher}, {Lee}, {Lin}, {Lin}, {Lunnan},
  {Giomi}, {Mahabal}, {Mao}, {Miller}, {Monkewitz}, {Murphy}, {Ngeow},
  {Nordin}, {Nugent}, {Ofek}, {Patterson}, {Penprase}, {Porter}, {Rauch},
  {Rebbapragada}, {Reiley}, {Rigault}, {Rodriguez}, {van Roestel}, {Rusholme},
  {van Santen}, {Schulze}, {Shupe}, {Singer}, {Soumagnac}, {Stein}, {Surace},
  {Sollerman}, {Szkody}, {Taddia}, {Terek}, {Van Sistine}, {van Velzen},
  {Vestrand}, {Walters}, {Ward}, {Ye}, {Yu}, {Yan}, \& {Zolkower}}]{Bellm2019}
{Bellm}, E.~C., {Kulkarni}, S.~R., {Graham}, M.~J., {et~al.} 2019, \pasp, 131,
  018002, \dodoi{10.1088/1538-3873/aaecbe}

\bibitem[{{Bovy}(2017)}]{Bovy2017}
{Bovy}, J. 2017, \mnras, 470, 1360, \dodoi{10.1093/mnras/stx1277}

\bibitem[{{Bovy} {et~al.}(2016){Bovy}, {Rix}, {Green}, {Schlafly}, \&
  {Finkbeiner}}]{Bovy2016}
{Bovy}, J., {Rix}, H.-W., {Green}, G.~M., {Schlafly}, E.~F., \& {Finkbeiner},
  D.~P. 2016, \apj, 818, 130, \dodoi{10.3847/0004-637X/818/2/130}

\bibitem[{{Bovy} {et~al.}(2012){Bovy}, {Rix}, \& {Hogg}}]{Bovy2012}
{Bovy}, J., {Rix}, H.-W., \& {Hogg}, D.~W. 2012, \apj, 751, 131,
  \dodoi{10.1088/0004-637X/751/2/131}

\bibitem[{{Boyer} {et~al.}(2019){Boyer}, {Williams}, {Aringer}, {Chen},
  {Dalcanton}, {Girardi}, {Guhathakurta}, {Marigo}, {Olsen}, {Rosenfield}, \&
  {Weisz}}]{Boyer2019}
{Boyer}, M.~L., {Williams}, B.~F., {Aringer}, B., {et~al.} 2019, \apj, 879,
  109, \dodoi{10.3847/1538-4357/ab24e2}

\bibitem[{{Busso} {et~al.}(1999){Busso}, {Gallino}, \&
  {Wasserburg}}]{Busso1999}
{Busso}, M., {Gallino}, R., \& {Wasserburg}, G.~J. 1999, \araa, 37, 239,
  \dodoi{10.1146/annurev.astro.37.1.239}

\bibitem[{{Canbay} {et~al.}(2023){Canbay}, {Bilir}, {{\"O}zd{\"o}nmez}, \&
  {Ak}}]{Canbay2023}
{Canbay}, R., {Bilir}, S., {{\"O}zd{\"o}nmez}, A., \& {Ak}, T. 2023, \aj, 165,
  163, \dodoi{10.3847/1538-3881/acbead}

\bibitem[{{Cardelli} {et~al.}(1989){Cardelli}, {Clayton}, \&
  {Mathis}}]{Cardelli1989}
{Cardelli}, J.~A., {Clayton}, G.~C., \& {Mathis}, J.~S. 1989, \apj, 345, 245,
  \dodoi{10.1086/167900}

\bibitem[{{Carrasco} {et~al.}(2021){Carrasco}, {Weiler}, {Jordi}, {Fabricius},
  {De Angeli}, {Evans}, {van Leeuwen}, {Riello}, \&
  {Montegriffo}}]{Carrasco2021}
{Carrasco}, J.~M., {Weiler}, M., {Jordi}, C., {et~al.} 2021, \aap, 652, A86,
  \dodoi{10.1051/0004-6361/202141249}

\bibitem[{{Chandra} {et~al.}(2023){Chandra}, {Naidu}, {Conroy}, {Ji}, {Rix},
  {Bonaca}, {Cargile}, {Han}, {Johnson}, {Ting}, {Woody}, \&
  {Zaritsky}}]{Chandra2023}
{Chandra}, V., {Naidu}, R.~P., {Conroy}, C., {et~al.} 2023, \apj, 951, 26,
  \dodoi{10.3847/1538-4357/accf13}

\bibitem[{{Chen} \& {Guestrin}(2016)}]{Chen2016}
{Chen}, T., \& {Guestrin}, C. 2016, arXiv e-prints, arXiv:1603.02754,
  \dodoi{10.48550/arXiv.1603.02754}

\bibitem[{{Christlieb} {et~al.}(2001){Christlieb}, {Green}, {Wisotzki}, \&
  {Reimers}}]{Christlieb2001}
{Christlieb}, N., {Green}, P.~J., {Wisotzki}, L., \& {Reimers}, D. 2001, \aap,
  375, 366, \dodoi{10.1051/0004-6361:20010814}

\bibitem[{{Claussen} {et~al.}(1987){Claussen}, {Kleinmann}, {Joyce}, \&
  {Jura}}]{Claussen1987}
{Claussen}, M.~J., {Kleinmann}, S.~G., {Joyce}, R.~R., \& {Jura}, M. 1987,
  \apjs, 65, 385, \dodoi{10.1086/191229}

\bibitem[{{Cristallo} {et~al.}(2016){Cristallo}, {Piersanti}, \&
  {Straniero}}]{Cristallo2016}
{Cristallo}, S., {Piersanti}, L., \& {Straniero}, O. 2016, in Journal of
  Physics Conference Series, Vol. 665, Journal of Physics Conference Series
  (IOP), 012019, \dodoi{10.1088/1742-6596/665/1/012019}

\bibitem[{{Cummings} {et~al.}(2018){Cummings}, {Kalirai}, {Tremblay},
  {Ramirez-Ruiz}, \& {Choi}}]{Cummings2018}
{Cummings}, J.~D., {Kalirai}, J.~S., {Tremblay}, P.~E., {Ramirez-Ruiz}, E., \&
  {Choi}, J. 2018, \apj, 866, 21, \dodoi{10.3847/1538-4357/aadfd6}

\bibitem[{{Dahn} {et~al.}(1977){Dahn}, {Liebert}, {Kron}, {Spinrad}, \&
  {Hintzen}}]{Dahn1977}
{Dahn}, C.~C., {Liebert}, J., {Kron}, R.~G., {Spinrad}, H., \& {Hintzen}, P.~M.
  1977, \apj, 216, 757, \dodoi{10.1086/155518}

\bibitem[{{Dalton} {et~al.}(2012){Dalton}, {Trager}, {Abrams}, {Carter},
  {Bonifacio}, {Aguerri}, {MacIntosh}, {Evans}, {Lewis}, {Navarro}, {Agocs},
  {Dee}, {Rousset}, {Tosh}, {Middleton}, {Pragt}, {Terrett}, {Brock}, {Benn},
  {Verheijen}, {Cano Infantes}, {Bevil}, {Steele}, {Mottram}, {Bates},
  {Gribbin}, {Rey}, {Rodriguez}, {Delgado}, {Guinouard}, {Walton}, {Irwin},
  {Jagourel}, {Stuik}, {Gerlofsma}, {Roelfsma}, {Skillen}, {Ridings},
  {Balcells}, {Daban}, {Gouvret}, {Venema}, \& {Girard}}]{WEAVE}
{Dalton}, G., {Trager}, S.~C., {Abrams}, D.~C., {et~al.} 2012, in Society of
  Photo-Optical Instrumentation Engineers (SPIE) Conference Series, Vol. 8446,
  Ground-based and Airborne Instrumentation for Astronomy IV, ed. I.~S.
  {McLean}, S.~K. {Ramsay}, \& H.~{Takami}, 84460P, \dodoi{10.1117/12.925950}

\bibitem[{{Dawson} {et~al.}(2024){Dawson}, {Geier}, {Heber}, {Pelisoli},
  {Dorsch}, {Schaffenroth}, {Reindl}, {Culpan}, {Pritzkuleit}, {Vos},
  {Soemitro}, {Roth}, {Schneider}, {Uzundag}, {Vu{\v{c}}kovi{\'c}}, {Antunes
  Amaral}, {Istrate}, {Justham}, {{\O}stensen}, {Telting}, {Djupvik}, {Raddi},
  {Green}, {Jeffery}, {Kepler}, {Munday}, {Steinmetz}, \&
  {Kupfer}}]{Dawson2024}
{Dawson}, H., {Geier}, S., {Heber}, U., {et~al.} 2024, \aap, 686, A25,
  \dodoi{10.1051/0004-6361/202348319}

\bibitem[{{De Angeli} {et~al.}(2023){De Angeli}, {Weiler}, {Montegriffo},
  {Evans}, {Riello}, {Andrae}, {Carrasco}, {Busso}, {Burgess}, {Cacciari},
  {Davidson}, {Harrison}, {Hodgkin}, {Jordi}, {Osborne}, {Pancino},
  {Altavilla}, {Barstow}, {Bailer-Jones}, {Bellazzini}, {Brown}, {Castellani},
  {Cowell}, {Delchambre}, {De Luise}, {Diener}, {Fabricius}, {Fouesneau},
  {Fr{\'e}mat}, {Gilmore}, {Giuffrida}, {Hambly}, {Hidalgo}, {Holland},
  {Kostrzewa-Rutkowska}, {van Leeuwen}, {Lobel}, {Marinoni}, {Miller},
  {Pagani}, {Palaversa}, {Piersimoni}, {Pulone}, {Ragaini}, {Rainer},
  {Richards}, {Rixon}, {Ruz-Mieres}, {Sanna}, {Sarro}, {Rowell}, {Sordo},
  {Walton}, \& {Yoldas}}]{DeAngeli2023}
{De Angeli}, F., {Weiler}, M., {Montegriffo}, P., {et~al.} 2023, \aap, 674, A2,
  \dodoi{10.1051/0004-6361/202243680}

\bibitem[{{de Jong} {et~al.}(2012){de Jong}, {Bellido-Tirado}, {Chiappini},
  {Depagne}, {Haynes}, {Johl}, {Schnurr}, {Schwope}, {Walcher}, {Dionies},
  {Haynes}, {Kelz}, {Kitaura}, {Lamer}, {Minchev}, {M{\"u}ller}, {Nuza},
  {Olaya}, {Piffl}, {Popow}, {Steinmetz}, {Ural}, {Williams}, {Winkler},
  {Wisotzki}, {Ansorge}, {Banerji}, {Gonzalez Solares}, {Irwin}, {Kennicutt},
  {King}, {McMahon}, {Koposov}, {Parry}, {Sun}, {Walton}, {Finger}, {Iwert},
  {Krumpe}, {Lizon}, {Vincenzo}, {Amans}, {Bonifacio}, {Cohen}, {Francois},
  {Jagourel}, {Mignot}, {Royer}, {Sartoretti}, {Bender}, {Grupp}, {Hess},
  {Lang-Bardl}, {Muschielok}, {B{\"o}hringer}, {Boller}, {Bongiorno}, {Brusa},
  {Dwelly}, {Merloni}, {Nandra}, {Salvato}, {Pragt}, {Navarro}, {Gerlofsma},
  {Roelfsema}, {Dalton}, {Middleton}, {Tosh}, {Boeche}, {Caffau}, {Christlieb},
  {Grebel}, {Hansen}, {Koch}, {Ludwig}, {Quirrenbach}, {Sbordone}, {Seifert},
  {Thimm}, {Trifonov}, {Helmi}, {Trager}, {Feltzing}, {Korn}, \&
  {Boland}}]{4MOST}
{de Jong}, R.~S., {Bellido-Tirado}, O., {Chiappini}, C., {et~al.} 2012, in
  Society of Photo-Optical Instrumentation Engineers (SPIE) Conference Series,
  Vol. 8446, Ground-based and Airborne Instrumentation for Astronomy IV, ed.
  I.~S. {McLean}, S.~K. {Ramsay}, \& H.~{Takami}, 84460T,
  \dodoi{10.1117/12.926239}

\bibitem[{{de Kool} \& {Green}(1995)}]{Kool1995}
{de Kool}, M., \& {Green}, P.~J. 1995, \apj, 449, 236, \dodoi{10.1086/176051}

\bibitem[{{Deng} {et~al.}(2012){Deng}, {Newberg}, {Liu}, {Carlin}, {Beers},
  {Chen}, {Chen}, {Christlieb}, {Grillmair}, {Guhathakurta}, {Han}, {Hou},
  {Lee}, {L{\'e}pine}, {Li}, {Liu}, {Pan}, {Sellwood}, {Wang}, {Wang}, {Yang},
  {Yanny}, {Zhang}, {Zhang}, {Zheng}, \& {Zhu}}]{Deng2012}
{Deng}, L.-C., {Newberg}, H.~J., {Liu}, C., {et~al.} 2012, Research in
  Astronomy and Astrophysics, 12, 735, \dodoi{10.1088/1674-4527/12/7/003}

\bibitem[{{DESI Collaboration} {et~al.}(2016){DESI Collaboration}, {Aghamousa},
  {Aguilar}, {Ahlen}, {Alam}, {Allen}, {Allende Prieto}, {Annis}, {Bailey},
  {Balland}, {Ballester}, {Baltay}, {Beaufore}, {Bebek}, {Beers}, {Bell},
  {Bernal}, {Besuner}, {Beutler}, {Blake}, {Bleuler}, {Blomqvist}, {Blum},
  {Bolton}, {Briceno}, {Brooks}, {Brownstein}, {Buckley-Geer}, {Burden},
  {Burtin}, {Busca}, {Cahn}, {Cai}, {Cardiel-Sas}, {Carlberg}, {Carton},
  {Casas}, {Castander}, {Cervantes-Cota}, {Claybaugh}, {Close}, {Coker},
  {Cole}, {Comparat}, {Cooper}, {Cousinou}, {Crocce}, {Cuby}, {Cunningham},
  {Davis}, {Dawson}, {de la Macorra}, {De Vicente}, {Delubac}, {Derwent},
  {Dey}, {Dhungana}, {Ding}, {Doel}, {Duan}, {Ealet}, {Edelstein},
  {Eftekharzadeh}, {Eisenstein}, {Elliott}, {Escoffier}, {Evatt}, {Fagrelius},
  {Fan}, {Fanning}, {Farahi}, {Farihi}, {Favole}, {Feng}, {Fernandez},
  {Findlay}, {Finkbeiner}, {Fitzpatrick}, {Flaugher}, {Flender}, {Font-Ribera},
  {Forero-Romero}, {Fosalba}, {Frenk}, {Fumagalli}, {Gaensicke}, {Gallo},
  {Garcia-Bellido}, {Gaztanaga}, {Pietro Gentile Fusillo}, {Gerard},
  {Gershkovich}, {Giannantonio}, {Gillet}, {Gonzalez-de-Rivera},
  {Gonzalez-Perez}, {Gott}, {Graur}, {Gutierrez}, {Guy}, {Habib}, {Heetderks},
  {Heetderks}, {Heitmann}, {Hellwing}, {Herrera}, {Ho}, {Holland}, {Honscheid},
  {Huff}, {Hutchinson}, {Huterer}, {Hwang}, {Illa Laguna}, {Ishikawa},
  {Jacobs}, {Jeffrey}, {Jelinsky}, {Jennings}, {Jiang}, {Jimenez}, {Johnson},
  {Joyce}, {Jullo}, {Juneau}, {Kama}, {Karcher}, {Karkar}, {Kehoe}, {Kennamer},
  {Kent}, {Kilbinger}, {Kim}, {Kirkby}, {Kisner}, {Kitanidis}, {Kneib},
  {Koposov}, {Kovacs}, {Koyama}, {Kremin}, {Kron}, {Kronig}, {Kueter-Young},
  {Lacey}, {Lafever}, {Lahav}, {Lambert}, {Lampton}, {Landriau}, {Lang},
  {Lauer}, {Le Goff}, {Le Guillou}, {Le Van Suu}, {Lee}, {Lee}, {Leitner},
  {Lesser}, {Levi}, {L'Huillier}, {Li}, {Liang}, {Lin}, {Linder}, {Loebman},
  {Luki{\'c}}, {Ma}, {MacCrann}, {Magneville}, {Makarem}, {Manera}, {Manser},
  {Marshall}, {Martini}, {Massey}, {Matheson}, {McCauley}, {McDonald},
  {McGreer}, {Meisner}, {Metcalfe}, {Miller}, {Miquel}, {Moustakas}, {Myers},
  {Naik}, {Newman}, {Nichol}, {Nicola}, {Nicolati da Costa}, {Nie}, {Niz},
  {Norberg}, {Nord}, {Norman}, {Nugent}, {O'Brien}, {Oh}, {Olsen}, {Padilla},
  {Padmanabhan}, {Padmanabhan}, {Palanque-Delabrouille}, {Palmese},
  {Pappalardo}, {P{\^a}ris}, {Park}, {Patej}, {Peacock}, {Peiris}, {Peng},
  {Percival}, {Perruchot}, {Pieri}, {Pogge}, {Pollack}, {Poppett}, {Prada},
  {Prakash}, {Probst}, {Rabinowitz}, {Raichoor}, {Ree}, {Refregier}, {Regal},
  {Reid}, {Reil}, {Rezaie}, {Rockosi}, {Roe}, {Ronayette}, {Roodman}, {Ross},
  {Ross}, {Rossi}, {Rozo}, {Ruhlmann-Kleider}, {Rykoff}, {Sabiu}, {Samushia},
  {Sanchez}, {Sanchez}, {Schlegel}, {Schneider}, {Schubnell}, {Secroun},
  {Seljak}, {Seo}, {Serrano}, {Shafieloo}, {Shan}, {Sharples}, {Sholl},
  {Shourt}, {Silber}, {Silva}, {Sirk}, {Slosar}, {Smith}, {Smoot}, {Som},
  {Song}, {Sprayberry}, {Staten}, {Stefanik}, {Tarle}, {Sien Tie}, {Tinker},
  {Tojeiro}, {Valdes}, {Valenzuela}, {Valluri}, {Vargas-Magana}, {Verde},
  {Walker}, {Wang}, {Wang}, {Weaver}, {Weaverdyck}, {Wechsler}, {Weinberg},
  {White}, {Yang}, {Yeche}, {Zhang}, {Zhao}, {Zheng}, {Zhou}, {Zhou}, {Zhu},
  {Zou}, \& {Zu}}]{DESI2016}
{DESI Collaboration}, {Aghamousa}, A., {Aguilar}, J., {et~al.} 2016, arXiv
  e-prints, arXiv:1611.00037, \dodoi{10.48550/arXiv.1611.00037}

\bibitem[{Dietterich(2000)}]{Dietterich2000}
Dietterich, T.~G. 2000, in Multiple Classifier Systems (Berlin, Heidelberg:
  Springer Berlin Heidelberg), 1--15

\bibitem[{{Drimmel} {et~al.}(2003){Drimmel}, {Cabrera-Lavers}, \&
  {L{\'o}pez-Corredoira}}]{Drimmel2003}
{Drimmel}, R., {Cabrera-Lavers}, A., \& {L{\'o}pez-Corredoira}, M. 2003, \aap,
  409, 205, \dodoi{10.1051/0004-6361:20031070}

\bibitem[{{Eyer} {et~al.}(2023){Eyer}, {Audard}, {Holl}, {Rimoldini},
  {Carnerero}, {Clementini}, {De Ridder}, {Distefano}, {Evans}, {Gavras},
  {Gomel}, {Lebzelter}, {Marton}, {Mowlavi}, {Panahi}, {Ripepi}, {Wyrzykowski},
  {Nienartowicz}, {Jevardat de Fombelle}, {Lecoeur-Taibi}, {Rohrbasser},
  {Riello}, {Garc{\'\i}a-Lario}, {Lanzafame}, {Mazeh}, {Raiteri}, {Zucker},
  {{\'A}brah{\'a}m}, {Aerts}, {Aguado}, {Anderson}, {Bashi}, {Binnenfeld},
  {Faigler}, {Garofalo}, {Karbevska}, {K{\'o}sp{\'a}l}, {Kruszy{\'n}ska},
  {Kun}, {Lanza}, {Leccia}, {Marconi}, {Messina}, {Molinaro}, {Moln{\'a}r},
  {Muraveva}, {Musella}, {Nagy}, {Pagano}, {Palaversa}, {Plachy}, {Pr{\v{s}}a},
  {Rybicki}, {Shahaf}, {Szabados}, {Szegedi-Elek}, {Trabucchi}, {Barblan},
  {Grenon}, {Roelens}, \& {S{\"u}veges}}]{GaiaDR3_multiepoch}
{Eyer}, L., {Audard}, M., {Holl}, B., {et~al.} 2023, \aap, 674, A13,
  \dodoi{10.1051/0004-6361/202244242}

\bibitem[{{Foreman-Mackey} {et~al.}(2013){Foreman-Mackey}, {Hogg}, {Lang}, \&
  {Goodman}}]{emcee}
{Foreman-Mackey}, D., {Hogg}, D.~W., {Lang}, D., \& {Goodman}, J. 2013, \pasp,
  125, 306, \dodoi{10.1086/670067}

\bibitem[{{Gaia Collaboration} {et~al.}(2023){Gaia Collaboration}, {Vallenari},
  {Brown}, {Prusti}, {de Bruijne}, {Arenou}, {Babusiaux}, {Biermann},
  {Creevey}, {Ducourant}, {Evans}, {Eyer}, {Guerra}, {Hutton}, {Jordi},
  {Klioner}, {Lammers}, {Lindegren}, {Luri}, {Mignard}, {Panem}, {Pourbaix},
  {Randich}, {Sartoretti}, {Soubiran}, {Tanga}, {Walton}, {Bailer-Jones},
  {Bastian}, {Drimmel}, {Jansen}, {Katz}, {Lattanzi}, {van Leeuwen}, {Bakker},
  {Cacciari}, {Casta{\~n}eda}, {De Angeli}, {Fabricius}, {Fouesneau},
  {Fr{\'e}mat}, {Galluccio}, {Guerrier}, {Heiter}, {Masana}, {Messineo},
  {Mowlavi}, {Nicolas}, {Nienartowicz}, {Pailler}, {Panuzzo}, {Riclet}, {Roux},
  {Seabroke}, {Sordo}, {Th{\'e}venin}, {Gracia-Abril}, {Portell}, {Teyssier},
  {Altmann}, {Andrae}, {Audard}, {Bellas-Velidis}, {Benson}, {Berthier},
  {Blomme}, {Burgess}, {Busonero}, {Busso}, {C{\'a}novas}, {Carry}, {Cellino},
  {Cheek}, {Clementini}, {Damerdji}, {Davidson}, {de Teodoro}, {Nu{\~n}ez
  Campos}, {Delchambre}, {Dell'Oro}, {Esquej}, {Fern{\'a}ndez-Hern{\'a}ndez},
  {Fraile}, {Garabato}, {Garc{\'\i}a-Lario}, {Gosset}, {Haigron}, {Halbwachs},
  {Hambly}, {Harrison}, {Hern{\'a}ndez}, {Hestroffer}, {Hodgkin}, {Holl},
  {Jan{\ss}en}, {Jevardat de Fombelle}, {Jordan}, {Krone-Martins}, {Lanzafame},
  {L{\"o}ffler}, {Marchal}, {Marrese}, {Moitinho}, {Muinonen}, {Osborne},
  {Pancino}, {Pauwels}, {Recio-Blanco}, {Reyl{\'e}}, {Riello}, {Rimoldini},
  {Roegiers}, {Rybizki}, {Sarro}, {Siopis}, {Smith}, {Sozzetti}, {Utrilla},
  {van Leeuwen}, {Abbas}, {{\'A}brah{\'a}m}, {Abreu Aramburu}, {Aerts},
  {Aguado}, {Ajaj}, {Aldea-Montero}, {Altavilla}, {{\'A}lvarez}, {Alves},
  {Anders}, {Anderson}, {Anglada Varela}, {Antoja}, {Baines}, {Baker},
  {Balaguer-N{\'u}{\~n}ez}, {Balbinot}, {Balog}, {Barache}, {Barbato},
  {Barros}, {Barstow}, {Bartolom{\'e}}, {Bassilana}, {Bauchet}, {Becciani},
  {Bellazzini}, {Berihuete}, {Bernet}, {Bertone}, {Bianchi}, {Binnenfeld},
  {Blanco-Cuaresma}, {Blazere}, {Boch}, {Bombrun}, {Bossini}, {Bouquillon},
  {Bragaglia}, {Bramante}, {Breedt}, {Bressan}, {Brouillet}, {Brugaletta},
  {Bucciarelli}, {Burlacu}, {Butkevich}, {Buzzi}, {Caffau}, {Cancelliere},
  {Cantat-Gaudin}, {Carballo}, {Carlucci}, {Carnerero}, {Carrasco},
  {Casamiquela}, {Castellani}, {Castro-Ginard}, {Chaoul}, {Charlot}, {Chemin},
  {Chiaramida}, {Chiavassa}, {Chornay}, {Comoretto}, {Contursi}, {Cooper},
  {Cornez}, {Cowell}, {Crifo}, {Cropper}, {Crosta}, {Crowley}, {Dafonte},
  {Dapergolas}, {David}, {David}, {de Laverny}, {De Luise}, \& {De
  March}}]{GaiaDR3}
{Gaia Collaboration}, {Vallenari}, A., {Brown}, A.~G.~A., {et~al.} 2023, \aap,
  674, A1, \dodoi{10.1051/0004-6361/202243940}

\bibitem[{{Garc{\'\i}a-Zamora} {et~al.}(2023){Garc{\'\i}a-Zamora}, {Torres}, \&
  {Rebassa-Mansergas}}]{Garcia-Zamora_2024}
{Garc{\'\i}a-Zamora}, E.~M., {Torres}, S., \& {Rebassa-Mansergas}, A. 2023,
  \aap, 679, A127, \dodoi{10.1051/0004-6361/202347601}

\bibitem[{{Gentile Fusillo} {et~al.}(2021){Gentile Fusillo}, {Tremblay},
  {Cukanovaite}, {Vorontseva}, {Lallement}, {Hollands}, {G{\"a}nsicke},
  {Burdge}, {McCleery}, \& {Jordan}}]{GentileFusillo2021}
{Gentile Fusillo}, N.~P., {Tremblay}, P.~E., {Cukanovaite}, E., {et~al.} 2021,
  \mnras, 508, 3877, \dodoi{10.1093/mnras/stab2672}

\bibitem[{{Gilmore} \& {Reid}(1983)}]{Gilmore1983}
{Gilmore}, G., \& {Reid}, N. 1983, \mnras, 202, 1025,
  \dodoi{10.1093/mnras/202.4.1025}

\bibitem[{{Girardi} \& {Marigo}(2007)}]{Girardi2007}
{Girardi}, L., \& {Marigo}, P. 2007, \aap, 462, 237,
  \dodoi{10.1051/0004-6361:20065249}

\bibitem[{{Goodman} \& {Weare}(2010)}]{Goodman2010}
{Goodman}, J., \& {Weare}, J. 2010, Communications in Applied Mathematics and
  Computational Science, 5, 65, \dodoi{10.2140/camcos.2010.5.65}

\bibitem[{{Gray}(2022)}]{Gray2022}
{Gray}, D.~F. 2022, {The Observation and Analysis of Stellar Photospheres}
  ({Cambridge University Press}), \dodoi{10.1017/9781009082136}

\bibitem[{{Green}(2013)}]{Green2013}
{Green}, P. 2013, \apj, 765, 12, \dodoi{10.1088/0004-637X/765/1/12}

\bibitem[{{Green} {et~al.}(1992){Green}, {Margon}, {Anderson}, \&
  {MacConnell}}]{Green1992}
{Green}, P.~J., {Margon}, B., {Anderson}, S.~F., \& {MacConnell}, D.~J. 1992,
  \apj, 400, 659, \dodoi{10.1086/172027}

\bibitem[{{Green} {et~al.}(1991){Green}, {Margon}, \& {MacConnell}}]{Green1991}
{Green}, P.~J., {Margon}, B., \& {MacConnell}, D.~J. 1991, \apjl, 380, L31,
  \dodoi{10.1086/186166}

\bibitem[{{Green} {et~al.}(2019){Green}, {Montez}, {Mazzoni}, {Filippazzo},
  {Anderson}, {De Marco}, {Drake}, {Farihi}, {Frank}, {Kastner}, {Miszalski},
  \& {Roulston}}]{Green2019}
{Green}, P.~J., {Montez}, R., {Mazzoni}, F., {et~al.} 2019, \apj, 881, 49,
  \dodoi{10.3847/1538-4357/ab2bf4}

\bibitem[{{Hansen} {et~al.}(2016){Hansen}, {Andersen}, {Nordstr{\"o}m},
  {Beers}, {Placco}, {Yoon}, \& {Buchhave}}]{Hansen2016}
{Hansen}, T.~T., {Andersen}, J., {Nordstr{\"o}m}, B., {et~al.} 2016, \aap, 588,
  A3, \dodoi{10.1051/0004-6361/201527409}

\bibitem[{Harris {et~al.}(2020)Harris, Millman, van~der Walt, Gommers,
  Virtanen, Cournapeau, Wieser, Taylor, Berg, Smith, Kern, Picus, Hoyer, van
  Kerkwijk, Brett, Haldane, Fernández~del Río, Wiebe, Peterson,
  Gérard-Marchant, Sheppard, Reddy, Weckesser, Abbasi, Gohlke, \&
  Oliphant}]{numpy}
Harris, C.~R., Millman, K.~J., van~der Walt, S.~J., {et~al.} 2020, Array
  programming with {NumPy}, \dodoi{10.1038/s41586-020-2649-2}

\bibitem[{{Haywood} {et~al.}(2013){Haywood}, {Di Matteo}, {Lehnert}, {Katz}, \&
  {G{\'o}mez}}]{Haywood2013}
{Haywood}, M., {Di Matteo}, P., {Lehnert}, M.~D., {Katz}, D., \& {G{\'o}mez},
  A. 2013, \aap, 560, A109, \dodoi{10.1051/0004-6361/201321397}

\bibitem[{He {et~al.}(2022)He, Luo, \& Chen}]{He2023}
He, X.-J., Luo, A.-L., \& Chen, Y.-Q. 2022, Monthly Notices of the Royal
  Astronomical Society, 512, 1710, \dodoi{10.1093/mnras/stac484}

\bibitem[{{Heber} {et~al.}(1993){Heber}, {Bade}, {Jordan}, \&
  {Voges}}]{Heber1993}
{Heber}, U., {Bade}, N., {Jordan}, S., \& {Voges}, W. 1993, \aap, 267, L31

\bibitem[{Hunter(2007)}]{matplotlib}
Hunter, J.~D. 2007, Computing in Science \& Engineering, 9, 90,
  \dodoi{10.1109/MCSE.2007.55}

\bibitem[{{Ibata} {et~al.}(2013){Ibata}, {Lewis}, {Martin}, {Bellazzini}, \&
  {Correnti}}]{Ibata2013}
{Ibata}, R., {Lewis}, G.~F., {Martin}, N.~F., {Bellazzini}, M., \& {Correnti},
  M. 2013, \apjl, 765, L15, \dodoi{10.1088/2041-8205/765/1/L15}

\bibitem[{{Iben}(1974)}]{Iben1974}
{Iben}, I., J. 1974, \araa, 12, 215,
  \dodoi{10.1146/annurev.aa.12.090174.001243}

\bibitem[{{Iben} \& {Renzini}(1983)}]{Iben1983}
{Iben}, I., J., \& {Renzini}, A. 1983, \araa, 21, 271,
  \dodoi{10.1146/annurev.aa.21.090183.001415}

\bibitem[{{Igoshev} {et~al.}(2020){Igoshev}, {Perets}, \&
  {Michaely}}]{Igoshev2020}
{Igoshev}, A.~P., {Perets}, H.~B., \& {Michaely}, E. 2020, \mnras, 494, 1448,
  \dodoi{10.1093/mnras/staa833}

\bibitem[{{Iwanek} {et~al.}(2021){Iwanek}, {Soszy{\'n}ski}, \&
  {Koz{\l}owski}}]{Iwanek2021}
{Iwanek}, P., {Soszy{\'n}ski}, I., \& {Koz{\l}owski}, S. 2021, \apj, 919, 99,
  \dodoi{10.3847/1538-4357/ac10c5}

\bibitem[{{Izzard} {et~al.}(2010){Izzard}, {Dermine}, \& {Church}}]{Izzard2010}
{Izzard}, R.~G., {Dermine}, T., \& {Church}, R.~P. 2010, \aap, 523, A10,
  \dodoi{10.1051/0004-6361/201015254}

\bibitem[{{Izzard} {et~al.}(2007){Izzard}, {Jeffery}, \&
  {Lattanzio}}]{Izzard2007}
{Izzard}, R.~G., {Jeffery}, C.~S., \& {Lattanzio}, J. 2007, in American
  Institute of Physics Conference Series, Vol. 948, Unsolved Problems in
  Stellar Physics: A Conference in Honor of Douglas Gough, ed. R.~J.
  {Stancliffe}, G.~{Houdek}, R.~G. {Martin}, \& C.~A. {Tout} (AIP), 51--55,
  \dodoi{10.1063/1.2819011}

\bibitem[{{Ji} {et~al.}(2016){Ji}, {Cui}, {Liu}, {Luo}, {Zhao}, \&
  {Zhang}}]{Ji2016}
{Ji}, W., {Cui}, W., {Liu}, C., {et~al.} 2016, \apjs, 226, 1,
  \dodoi{10.3847/0067-0049/226/1/1}

\bibitem[{{Jia} {et~al.}(2023){Jia}, {Guo}, {Zhu}, {Li}, {Ma}, \&
  {L{\"u}}}]{Jia2023}
{Jia}, Y., {Guo}, S., {Zhu}, C., {et~al.} 2023, Research in Astronomy and
  Astrophysics, 23, 105012, \dodoi{10.1088/1674-4527/ace9b2}

\bibitem[{{Jorissen} {et~al.}(1998){Jorissen}, {Van Eck}, {Mayor}, \&
  {Udry}}]{Jorissen1998}
{Jorissen}, A., {Van Eck}, S., {Mayor}, M., \& {Udry}, S. 1998, \aap, 332, 877,
  \dodoi{10.48550/arXiv.astro-ph/9801272}

\bibitem[{{Jorissen} {et~al.}(2016){Jorissen}, {Van Eck}, {Van Winckel},
  {Merle}, {Boffin}, {Andersen}, {Nordstr{\"o}m}, {Udry}, {Masseron},
  {Lenaerts}, \& {Waelkens}}]{Jorissen2016}
{Jorissen}, A., {Van Eck}, S., {Van Winckel}, H., {et~al.} 2016, \aap, 586,
  A158, \dodoi{10.1051/0004-6361/201526992}

\bibitem[{{Kalirai} {et~al.}(2014){Kalirai}, {Marigo}, \&
  {Tremblay}}]{Kalirai2014}
{Kalirai}, J.~S., {Marigo}, P., \& {Tremblay}, P.-E. 2014, \apj, 782, 17,
  \dodoi{10.1088/0004-637X/782/1/17}

\bibitem[{{Karakas} \& {Lattanzio}(2014)}]{Karakas2014}
{Karakas}, A.~I., \& {Lattanzio}, J.~C. 2014, \pasa, 31, e030,
  \dodoi{10.1017/pasa.2014.21}

\bibitem[{{Kollmeier} {et~al.}(2017){Kollmeier}, {Zasowski}, {Rix}, {Johns},
  {Anderson}, {Drory}, {Johnson}, {Pogge}, {Bird}, {Blanc}, {Brownstein},
  {Crane}, {De Lee}, {Klaene}, {Kreckel}, {MacDonald}, {Merloni}, {Ness},
  {O'Brien}, {Sanchez-Gallego}, {Sayres}, {Shen}, {Thakar}, {Tkachenko},
  {Aerts}, {Blanton}, {Eisenstein}, {Holtzman}, {Maoz}, {Nandra}, {Rockosi},
  {Weinberg}, {Bovy}, {Casey}, {Chaname}, {Clerc}, {Conroy}, {Eracleous},
  {G{\"a}nsicke}, {Hekker}, {Horne}, {Kauffmann}, {McQuinn}, {Pellegrini},
  {Schinnerer}, {Schlafly}, {Schwope}, {Seibert}, {Teske}, \& {van
  Saders}}]{SDSS_5}
{Kollmeier}, J.~A., {Zasowski}, G., {Rix}, H.-W., {et~al.} 2017, arXiv
  e-prints, arXiv:1711.03234.
\newblock \doarXiv{1711.03234}

\bibitem[{{Kroupa}(2001)}]{Kroupa2001}
{Kroupa}, P. 2001, \mnras, 322, 231, \dodoi{10.1046/j.1365-8711.2001.04022.x}

\bibitem[{{Lebzelter} {et~al.}(2018){Lebzelter}, {Mowlavi}, {Marigo},
  {Pastorelli}, {Trabucchi}, {Wood}, \& {Lecoeur-Ta{\"\i}bi}}]{Lebzelter2018}
{Lebzelter}, T., {Mowlavi}, N., {Marigo}, P., {et~al.} 2018, \aap, 616, L13,
  \dodoi{10.1051/0004-6361/201833615}

\bibitem[{Li {et~al.}(2021)Li, Zhang, Cui, Fan, Zhao, Wu, He, Xu, Li, Han, Tao,
  Mi, Yang, \& Yang}]{Li2021}
Li, C., Zhang, Y., Cui, C., {et~al.} 2021, Monthly Notices of the Royal
  Astronomical Society, 506, 1651, \dodoi{10.1093/mnras/stab1650}

\bibitem[{{Li} {et~al.}(2024){Li}, {Zhang}, {Cui}, {Shi}, {Ji}, {Huo}, {Gao},
  {Zhang}, \& {Sun}}]{Li2024}
{Li}, L., {Zhang}, K., {Cui}, W., {et~al.} 2024, \apjs, 271, 12,
  \dodoi{10.3847/1538-4365/ad1881}

\bibitem[{{Li} {et~al.}(2018){Li}, {Luo}, {Du}, {Zuo}, {Wang}, {Zhao}, {Jiang},
  {Zhang}, {Liu}, {Qin}, {Wang}, {Du}, {Guo}, {Wang}, {Han}, {Xiang}, {Huang},
  {Chen}, {Chen}, {Kong}, {Hou}, {Song}, {Wang}, {Wu}, {Zhang}, {Zhang},
  {Wang}, {Cao}, {Hou}, \& {Zhao}}]{Li2018}
{Li}, Y.-B., {Luo}, A.~L., {Du}, C.-D., {et~al.} 2018, \apjs, 234, 31,
  \dodoi{10.3847/1538-4365/aaa415}

\bibitem[{{Liebert} {et~al.}(1994){Liebert}, {Schmidt}, {Lesser}, {Stepanian},
  {Lipovetsky}, {Chaffe}, {Foltz}, \& {Bergeron}}]{Liebert1994}
{Liebert}, J., {Schmidt}, G.~D., {Lesser}, M., {et~al.} 1994, \apj, 421, 733,
  \dodoi{10.1086/173685}

\bibitem[{{Lucatello} {et~al.}(2005){Lucatello}, {Tsangarides}, {Beers},
  {Carretta}, {Gratton}, \& {Ryan}}]{Lucatello2005}
{Lucatello}, S., {Tsangarides}, S., {Beers}, T.~C., {et~al.} 2005, \apj, 625,
  825, \dodoi{10.1086/428104}

\bibitem[{{Lucey} {et~al.}(2023){Lucey}, {Al Kharusi}, {Hawkins}, {Ting},
  {Ramachandra}, {Price-Whelan}, {Beers}, {Lee}, \& {Yoon}}]{Lucey2023}
{Lucey}, M., {Al Kharusi}, N., {Hawkins}, K., {et~al.} 2023, \mnras, 523, 4049,
  \dodoi{10.1093/mnras/stad1675}

\bibitem[{{Marshall} {et~al.}(2006){Marshall}, {Robin}, {Reyl{\'e}},
  {Schultheis}, \& {Picaud}}]{Marshall2006}
{Marshall}, D.~J., {Robin}, A.~C., {Reyl{\'e}}, C., {Schultheis}, M., \&
  {Picaud}, S. 2006, \aap, 453, 635, \dodoi{10.1051/0004-6361:20053842}

\bibitem[{{McClure} \& {Woodsworth}(1990)}]{McClure1990}
{McClure}, R.~D., \& {Woodsworth}, A.~W. 1990, \apj, 352, 709,
  \dodoi{10.1086/168573}

\bibitem[{{Nordstr{\"o}m} {et~al.}(2004){Nordstr{\"o}m}, {Mayor}, {Andersen},
  {Holmberg}, {Pont}, {J{\o}rgensen}, {Olsen}, {Udry}, \&
  {Mowlavi}}]{Nordstrom2004}
{Nordstr{\"o}m}, B., {Mayor}, M., {Andersen}, J., {et~al.} 2004, \aap, 418,
  989, \dodoi{10.1051/0004-6361:20035959}

\bibitem[{{O'Brien} {et~al.}(2024){O'Brien}, {Tremblay}, {Klein}, {Koester},
  {Melis}, {B{\'e}dard}, {Cukanovaite}, {Cunningham}, {Doyle}, {G{\"a}nsicke},
  {Gentile Fusillo}, {Hollands}, {McCleery}, {Pelisoli}, {Toonen},
  {Weinberger}, \& {Zuckerman}}]{OBrien2024}
{O'Brien}, M.~W., {Tremblay}, P.~E., {Klein}, B.~L., {et~al.} 2024, \mnras,
  527, 8687, \dodoi{10.1093/mnras/stad3773}

\bibitem[{{Ochsenbein} {et~al.}(2000){Ochsenbein}, {Bauer}, \&
  {Marcout}}]{Vizier2000}
{Ochsenbein}, F., {Bauer}, P., \& {Marcout}, J. 2000, \aaps, 143, 23,
  \dodoi{10.1051/aas:2000169}

\bibitem[{{Pala} {et~al.}(2020){Pala}, {G{\"a}nsicke}, {Breedt}, {Knigge},
  {Hermes}, {Gentile Fusillo}, {Hollands}, {Naylor}, {Pelisoli}, {Schreiber},
  {Toonen}, {Aungwerojwit}, {Cukanovaite}, {Dennihy}, {Manser}, {Pretorius},
  {Scaringi}, \& {Toloza}}]{Pala2020}
{Pala}, A.~F., {G{\"a}nsicke}, B.~T., {Breedt}, E., {et~al.} 2020, \mnras, 494,
  3799, \dodoi{10.1093/mnras/staa764}

\bibitem[{{Pastorelli} {et~al.}(2020){Pastorelli}, {Marigo}, {Girardi},
  {Aringer}, {Chen}, {Rubele}, {Trabucchi}, {Bladh}, {Boyer}, {Bressan},
  {Dalcanton}, {Groenewegen}, {Lebzelter}, {Mowlavi}, {Chubb}, {Cioni}, {de
  Grijs}, {Ivanov}, {Nanni}, {van Loon}, \& {Zaggia}}]{Pastorelli2020}
{Pastorelli}, G., {Marigo}, P., {Girardi}, L., {et~al.} 2020, \mnras, 498,
  3283, \dodoi{10.1093/mnras/staa2565}

\bibitem[{{Pecaut} \& {Mamajek}(2013)}]{Pecaut2013}
{Pecaut}, M.~J., \& {Mamajek}, E.~E. 2013, \apjs, 208, 9,
  \dodoi{10.1088/0067-0049/208/1/9}

\bibitem[{Pedregosa {et~al.}(2011)Pedregosa, Varoquaux, Gramfort, Michel,
  Thirion, Grisel, Blondel, Prettenhofer, Weiss, Dubourg, Vanderplas, Passos,
  Cournapeau, Brucher, Perrot, \& Duchesnay}]{Scikit-learn}
Pedregosa, F., Varoquaux, G., Gramfort, A., {et~al.} 2011, Journal of Machine
  Learning Research, 12, 2825

\bibitem[{{Plez} \& {Cohen}(2005)}]{Plez2005}
{Plez}, B., \& {Cohen}, J.~G. 2005, \aap, 434, 1117,
  \dodoi{10.1051/0004-6361:20042082}

\bibitem[{{Rebassa-Mansergas} {et~al.}(2021){Rebassa-Mansergas}, {Solano},
  {Jim{\'e}nez-Esteban}, {Torres}, {Rodrigo}, {Ferrer-Burjachs}, {Calcaferro},
  {Althaus}, \& {C{\'o}rsico}}]{Rebassa-Mansergas2021}
{Rebassa-Mansergas}, A., {Solano}, E., {Jim{\'e}nez-Esteban}, F.~M., {et~al.}
  2021, \mnras, 506, 5201, \dodoi{10.1093/mnras/stab2039}

\bibitem[{{Ricker} {et~al.}(2015){Ricker}, {Winn}, {Vanderspek}, {Latham},
  {Bakos}, {Bean}, {Berta-Thompson}, {Brown}, {Buchhave}, {Butler}, {Butler},
  {Chaplin}, {Charbonneau}, {Christensen-Dalsgaard}, {Clampin}, {Deming},
  {Doty}, {De Lee}, {Dressing}, {Dunham}, {Endl}, {Fressin}, {Ge}, {Henning},
  {Holman}, {Howard}, {Ida}, {Jenkins}, {Jernigan}, {Johnson}, {Kaltenegger},
  {Kawai}, {Kjeldsen}, {Laughlin}, {Levine}, {Lin}, {Lissauer}, {MacQueen},
  {Marcy}, {McCullough}, {Morton}, {Narita}, {Paegert}, {Palle}, {Pepe},
  {Pepper}, {Quirrenbach}, {Rinehart}, {Sasselov}, {Sato}, {Seager},
  {Sozzetti}, {Stassun}, {Sullivan}, {Szentgyorgyi}, {Torres}, {Udry}, \&
  {Villasenor}}]{Ricker2015}
{Ricker}, G.~R., {Winn}, J.~N., {Vanderspek}, R., {et~al.} 2015, Journal of
  Astronomical Telescopes, Instruments, and Systems, 1, 014003,
  \dodoi{10.1117/1.JATIS.1.1.014003}

\bibitem[{{Rix} {et~al.}(2021){Rix}, {Hogg}, {Boubert}, {Brown}, {Casey},
  {Drimmel}, {Everall}, {Fouesneau}, \& {Price-Whelan}}]{Rix2021}
{Rix}, H.-W., {Hogg}, D.~W., {Boubert}, D., {et~al.} 2021, \aj, 162, 142,
  \dodoi{10.3847/1538-3881/ac0c13}

\bibitem[{{Roulston} {et~al.}(2020){Roulston}, {Green}, \&
  {Kesseli}}]{Roulston2020}
{Roulston}, B.~R., {Green}, P.~J., \& {Kesseli}, A.~Y. 2020, \apjs, 249, 34,
  \dodoi{10.3847/1538-4365/aba1e7}

\bibitem[{{Roulston} {et~al.}(2021){Roulston}, {Green}, {Toonen}, \&
  {Hermes}}]{Roulston2021}
{Roulston}, B.~R., {Green}, P.~J., {Toonen}, S., \& {Hermes}, J.~J. 2021, \apj,
  922, 33, \dodoi{10.3847/1538-4357/ac157c}

\bibitem[{{Roulston} {et~al.}(2019){Roulston}, {Green}, {Ruan}, {MacLeod},
  {Anderson}, {Badenes}, {Brownstein}, {Schneider}, \&
  {Stassun}}]{Roulston2019}
{Roulston}, B.~R., {Green}, P.~J., {Ruan}, J.~J., {et~al.} 2019, \apj, 877, 44,
  \dodoi{10.3847/1538-4357/ab1a3e}

\bibitem[{{Roulston} {et~al.}(2022){Roulston}, {Green}, {Montez}, {Filippazzo},
  {Drake}, {Toonen}, {Anderson}, {Eracleous}, \& {Frank}}]{Roulston2022}
{Roulston}, B.~R., {Green}, P.~J., {Montez}, R., {et~al.} 2022, \apj, 926, 210,
  \dodoi{10.3847/1538-4357/ac4706}

\bibitem[{{Ruz-Mieres} \& {zuzannakr}(2024)}]{GaiaXPy}
{Ruz-Mieres}, D., \& {zuzannakr}. 2024, {gaia-dpci/GaiaXPy: GaiaXPy v2.1.2},
  \dodoi{10.5281/zenodo.11617977}

\bibitem[{{Sanduleak} \& {Pesch}(1988)}]{Sanduleak1988}
{Sanduleak}, N., \& {Pesch}, P. 1988, \apjs, 66, 387, \dodoi{10.1086/191262}

\bibitem[{{Science Software Branch at STScI}(2012)}]{Pyraf}
{Science Software Branch at STScI}. 2012, {PyRAF: Python alternative for IRAF}

\bibitem[{{Si} {et~al.}(2014){Si}, {Luo}, {Li}, {Zhang}, {Wei}, {Wu}, {Wu}, \&
  {Zhao}}]{Si2014}
{Si}, J., {Luo}, A., {Li}, Y., {et~al.} 2014, Science China Physics, Mechanics,
  and Astronomy, 57, 176, \dodoi{10.1007/s11433-013-5374-0}

\bibitem[{{Si} {et~al.}(2015){Si}, {Li}, {Luo}, {Tu}, {Shi}, {Zhang}, {Wei},
  {Zhao}, {Wu}, {Wu}, \& {Zhao}}]{Si2015}
{Si}, J.-M., {Li}, Y.-B., {Luo}, A.~L., {et~al.} 2015, Research in Astronomy
  and Astrophysics, 15, 1671, \dodoi{10.1088/1674-4527/15/10/005}

\bibitem[{{Sithajan} \& {Meethong}(2023)}]{Sithajan2023}
{Sithajan}, S., \& {Meethong}, S. 2023, \pasp, 135, 044502,
  \dodoi{10.1088/1538-3873/acc974}

\bibitem[{Solorio-Ramírez {et~al.}(2023)Solorio-Ramírez, Jiménez-Cruz,
  Villuendas-Rey, \& Yáñez-Márquez}]{Solorio-ramirez2023}
Solorio-Ramírez, J.-L., Jiménez-Cruz, R., Villuendas-Rey, Y., \&
  Yáñez-Márquez, C. 2023, Algorithms, 16, 293, \dodoi{10.3390/a16060293}

\bibitem[{{Stancliffe} {et~al.}(2005){Stancliffe}, {Izzard}, \&
  {Tout}}]{Stancliffe2005}
{Stancliffe}, R.~J., {Izzard}, R.~G., \& {Tout}, C.~A. 2005, \mnras, 356, L1,
  \dodoi{10.1111/j.1745-3933.2005.08491.x}

\bibitem[{{Stephenson}(1985)}]{Stephenson1985}
{Stephenson}, C.~B. 1985, \aj, 90, 784, \dodoi{10.1086/113787}

\bibitem[{{Straniero} {et~al.}(2023){Straniero}, {Abia}, \&
  {Dom{\'\i}nguez}}]{Straniero2023}
{Straniero}, O., {Abia}, C., \& {Dom{\'\i}nguez}, I. 2023, European Physical
  Journal A, 59, 17, \dodoi{10.1140/epja/s10050-023-00926-8}

\bibitem[{{Taylor}(2005)}]{topcat}
{Taylor}, M.~B. 2005, in Astronomical Society of the Pacific Conference Series,
  Vol. 347, Astronomical Data Analysis Software and Systems XIV, ed.
  P.~{Shopbell}, M.~{Britton}, \& R.~{Ebert}, 29

\bibitem[{{Tody}(1986)}]{Tody1986}
{Tody}, D. 1986, in Society of Photo-Optical Instrumentation Engineers (SPIE)
  Conference Series, Vol. 627, Instrumentation in astronomy VI, ed. D.~L.
  {Crawford}, 733, \dodoi{10.1117/12.968154}

\bibitem[{{Tody}(1993)}]{Tody1993}
---. 1993, in Astronomical Society of the Pacific Conference Series, Vol.~52,
  Astronomical Data Analysis Software and Systems II, ed. R.~J. {Hanisch},
  R.~J.~V. {Brissenden}, \& J.~{Barnes}, 173

\bibitem[{{Tonry} {et~al.}(2018){Tonry}, {Denneau}, {Flewelling}, {Heinze},
  {Onken}, {Smartt}, {Stalder}, {Weiland}, \& {Wolf}}]{Tonry2018}
{Tonry}, J.~L., {Denneau}, L., {Flewelling}, H., {et~al.} 2018, \apj, 867, 105,
  \dodoi{10.3847/1538-4357/aae386}

\bibitem[{{Totten} \& {Irwin}(1998)}]{Totten1998}
{Totten}, E.~J., \& {Irwin}, M.~J. 1998, \mnras, 294, 1,
  \dodoi{10.1046/j.1365-8711.1998.01086.x}

\bibitem[{{{\v{C}}otar} {et~al.}(2019){{\v{C}}otar}, {Zwitter}, {Kos},
  {Munari}, {Martell}, {Asplund}, {Bland-Hawthorn}, {Buder}, {de Silva},
  {Freeman}, {Sharma}, {Anguiano}, {Carollo}, {Horner}, {Lewis}, {Nataf},
  {Nordlander}, {Stello}, {Ting}, {Tinney}, {Traven}, {Wittenmyer}, \& {Galah
  Collaboration}}]{Cotar2019}
{{\v{C}}otar}, K., {Zwitter}, T., {Kos}, J., {et~al.} 2019, \mnras, 483, 3196,
  \dodoi{10.1093/mnras/sty3155}

\bibitem[{{Ventura} {et~al.}(2020){Ventura}, {Dell'Agli}, {Lugaro}, {Romano},
  {Tailo}, \& {Yag{\"u}e}}]{Ventura2020}
{Ventura}, P., {Dell'Agli}, F., {Lugaro}, M., {et~al.} 2020, \aap, 641, A103,
  \dodoi{10.1051/0004-6361/202038289}

\bibitem[{Virtanen {et~al.}(2020)Virtanen, Gommers, Oliphant, Haberland, Reddy,
  Cournapeau, Burovski, Peterson, Weckesser, Bright, {van der Walt}, Brett,
  Wilson, Millman, Mayorov, Nelson, Jones, Kern, Larson, Carey, Polat, Feng,
  Moore, {VanderPlas}, Laxalde, Perktold, Cimrman, Henriksen, Quintero, Harris,
  Archibald, Ribeiro, Pedregosa, {van Mulbregt}, \& {SciPy 1.0
  Contributors}}]{scipy}
Virtanen, P., Gommers, R., Oliphant, T.~E., {et~al.} 2020, Nature Methods, 17,
  261, \dodoi{10.1038/s41592-019-0686-2}

\bibitem[{{Wallerstein} \& {Knapp}(1998)}]{Wallerstein1998}
{Wallerstein}, G., \& {Knapp}, G.~R. 1998, \araa, 36, 369,
  \dodoi{10.1146/annurev.astro.36.1.369}

\bibitem[{{Wenger} {et~al.}(2000){Wenger}, {Ochsenbein}, {Egret}, {Dubois},
  {Bonnarel}, {Borde}, {Genova}, {Jasniewicz}, {Lalo{\"e}}, {Lesteven}, \&
  {Monier}}]{Simbad2000}
{Wenger}, M., {Ochsenbein}, F., {Egret}, D., {et~al.} 2000, \aaps, 143, 9,
  \dodoi{10.1051/aas:2000332}

\bibitem[{{Whitehouse} {et~al.}(2018){Whitehouse}, {Farihi}, {Green}, {Wilson},
  \& {Subasavage}}]{Whitehouse2018}
{Whitehouse}, L.~J., {Farihi}, J., {Green}, P.~J., {Wilson}, T.~G., \&
  {Subasavage}, J.~P. 2018, \mnras, 479, 3873, \dodoi{10.1093/mnras/sty1622}

\bibitem[{{Yao} {et~al.}(2023){Yao}, {Ji}, {Koposov}, \& {Limberg}}]{Yao2023}
{Yao}, Y., {Ji}, A.~P., {Koposov}, S.~E., \& {Limberg}, G. 2023, Monthly
  Notices of the Royal Astronomical Society, 527, 10937,
  \dodoi{10.1093/mnras/stad3775}

\bibitem[{{Yi} {et~al.}(2019){Yi}, {Chen}, {Pan}, {Yue}, {Lu}, {Li}, \&
  {Luo}}]{Yi2019}
{Yi}, Z., {Chen}, Z., {Pan}, J., {et~al.} 2019, \apj, 887, 241,
  \dodoi{10.3847/1538-4357/ab54d0}

\bibitem[{{Yue} {et~al.}(2021){Yue}, {Yi}, {Pan}, {Li}, \& {Li}}]{Yue2021}
{Yue}, L., {Yi}, Z., {Pan}, J., {Li}, X., \& {Li}, J. 2021, Optik, 225, 165535,
  \dodoi{10.1016/j.ijleo.2020.165535}

\bibitem[{{Zamora} {et~al.}(2009){Zamora}, {Abia}, {Plez}, {Dom{\'\i}nguez}, \&
  {Cristallo}}]{Zamora2009}
{Zamora}, O., {Abia}, C., {Plez}, B., {Dom{\'\i}nguez}, I., \& {Cristallo}, S.
  2009, \aap, 508, 909, \dodoi{10.1051/0004-6361/200912843}

\bibitem[{{Zhang} {et~al.}(2020){Zhang}, {Zhang}, \& {Zhao}}]{Zhang2020}
{Zhang}, J., {Zhang}, Y., \& {Zhao}, Y. 2020, \apjs, 246, 8,
  \dodoi{10.3847/1538-4365/ab5a7c}

\bibitem[{{Zhao} {et~al.}(2012){Zhao}, {Zhao}, {Chu}, {Jing}, \&
  {Deng}}]{Zhao2012}
{Zhao}, G., {Zhao}, Y.-H., {Chu}, Y.-Q., {Jing}, Y.-P., \& {Deng}, L.-C. 2012,
  Research in Astronomy and Astrophysics, 12, 723,
  \dodoi{10.1088/1674-4527/12/7/002}

\end{thebibliography}
\bibliographystyle{aasjournal}

\end{document}